\documentclass[aps,prb,twocolumn,floatfix,floats,showpacs,superscriptaddress,raggedbottom]{revtex4}
\usepackage{graphicx,latexsym}
\usepackage{dcolumn}
\usepackage{amsmath}
\pdfoutput=1

\begin{document}

\title{Transient magnetotransport through a quantum wire}

\author{Vidar Gudmundsson}
\email{vidar@raunvis.hi.is}
\affiliation{Science Institute, University of Iceland, Dunhaga 3,
        IS-107 Reykjavik, Iceland}
\affiliation{Physics Division, National Center for Theoretical
        Sciences, P.O.\ Box 2-131, Hsinchu 30013, Taiwan}
\author{Gunnar Thorgilsson}
\affiliation{Science Institute, University of Iceland, Dunhaga 3,
        IS-107 Reykjavik, Iceland}
\author{Chi-Shung Tang}
\email{cstang@nuu.edu.tw}
\affiliation{Research Center for Applied Sciences, Academia Sinica,
        Taipei 11529, Taiwan}
\affiliation{Physics Division, National Center for Theoretical
        Sciences, P.O.\ Box 2-131, Hsinchu 30013, Taiwan}
\affiliation{Department of Mechanical Engineering 
        National United University,
        1, Lienda, Miaoli 36003, Taiwan}

\author{Valeriu Moldoveanu}
\affiliation{National Institute of Materials Physics, P.O. Box MG-7,
Bucharest-Magurele, Romania}

%

\begin{abstract}
We consider an ideal parabolic quantum wire in a perpendicular magnetic field.
A simple Gaussian shaped scattering potential well or hill is flashed
softly on and off with its maximum at $t=0$, mimicking
a temporary broadening or narrowing of the wire. 
By an extension of the Lippmann-Schwinger formalism to time-dependent
scattering potentials we investigate the effects
on the continuous current that is driven through the
quantum wire with a vanishingly small forward bias. The Lippmann-Schwinger
approach to the scattering process enables us to investigate the interplay
between geometrical effects and effects caused by the magnetic field. 
\end{abstract}



\maketitle

%
%

\section{Introduction}

The confining potentials that define quantum dot systems or quantum wires set also 
their transport properties. For example, increasing the coupling between a quantum dot 
and a one-channel lead one can switch between the Coulomb blockade and mesoscopic 
Fano effect. \cite{Johnson:04:106803} Also, quantum dot arrays are formed by metallic gates suitably
arranged. \cite{Wiel02:1} In recent years it has become clear that time-dependent
potentials that are applied locally modify the shape of a mesoscopic system, and this in turn
drastically changes its transport properties. Typical examples are adiabatic quantum 
pumps,\cite{Switkes99:1905} i.e.\ unbiased systems that still allow charge transfer when slowly oscillating
potentials are applied to different regions. 

A crucial issue in time-dependent transport phenomena concerns transient current which
at a deeper level actually demands understanding of the fate of wave packets that propagate
through semiconductor devices driven by pulses having specific shapes.
This is a mandatory task in order to do signal processing in quantum computation.

From the theoretical point of view the transient regime in electronic
transport through nanostructures submitted to time-dependent potentials has
recently been studied using several methods. As we shall briefly describe below,
each of these approaches describes transients that are due to different perturbations.
Kurth {\it et al.}\ established a tractable scheme for computing transient
currents through one-dimensional quantum wires submitted to a time-dependent
bias.\cite{Kurth05:035308} Their method relies
on the Crank-Nicholson algorithm and on DFT calculations.
In our recent work\cite{Moldoveanu07:085330}
we investigated, within the Keldysh formalism, the transient currents that appear
when many-level quantum dots are suddenly coupled to leads on which a
finite (constant) bias is applied. Also, nonadiabatic pumping
transport has been studied.\cite{Moldoveanu07:0706.0968}
The two approaches are rather complimentary, in the sense that Kurth {\it et al.}\ 
were able to present transport properties in the presence of a time-dependent bias applied
while in Refs.\ \onlinecite{Moldoveanu07:085330} and \onlinecite{Moldoveanu07:0706.0968}
we showed transient behavior as well as the passage
to steady-state regime for systems submitted to finite bias and time-dependent signals
{\it at the contacts}. Otherwise stated,
in Refs.\ \onlinecite{Moldoveanu07:085330} and \onlinecite{Moldoveanu07:0706.0968} the transients appear
because the tunneling barriers between the leads and the system are time-dependent.
Experimental realizations using this driving mechanism include pump-and-probe 
techniques\cite{Fujisawa03:R1395} and turnstile 
pumps\cite{Kouwenhoven91:1991} and are expected to play an important role in qubit manipulation.
Another interesting problem is the pulse propagation along a quantum wire with embedded dots,
for which a scattering approach was recently developed by
Thorgilsson {\it et al.},\cite{Thorgilsson07:0708.0103} and Szafran and
Peeters.\cite{Szafran05:165301}

In this work we add another scenario for transient transport, namely we discuss
the transmission properties of a parabolic quantum wire when a space and time-dependent
potential is established on a finite region of the wire. The effect of such a potential is
to change (both {\it locally } and {\it dynamically}) the shape of the wire. Therefore one
has to study transport in the presence of time-dependent scattering.
This is a different and complimentary problem with respect to the
previous approaches,\cite{Kurth05:035308,Moldoveanu07:085330,Moldoveanu07:0706.0968}
in the sense that here the time-dependent
perturbation is applied neither on leads nor at the contacts but {\it on} the system itself.
Moreover, we consider a spatial dependence of the perturbation as well.

We believe that the problem we consider here is interesting for at least two reasons.
First, the modulation of the current in quantum wires by applying suitable time-dependent signals is
clearly possible in real life experiments, hence theoretical predictions on the
transport properties of such systems are important. Secondly, in order to study the
problem at hand we develop a time-dependent scattering framework for non-periodic
potentials which generalizes in some sense both the Floquet or adiabatic scattering
approaches that were extensively used in the context of quantum pumping
(see the review by L.\ Arrachea and M.\ Moskalets\cite{Arrachea06:245322}).
Earlier work built on wave function matching considering quantum transport with periodic 
time-dependent potentials acting on the system has been published by Tang and 
Chu.\cite{Tang96:4838,Tang99:1830,Tang01:353,Chung04:085315}
Switching properties of a T-shaped quantum waveguide have been studied by solving 
numerically directly the two-dimensional time-dependent Schr{\"o}dinger equation under 
continuous injection by Burgnies {\it et al.}\  as the stublength is changed without an 
external magnetic field.\cite{Burgnies97:803} They observe time-delays and mode-mixing 
that can be compared to the results of our calculation.

Our present model is strictly a one-electron model to observe the time-dependent 
interplay between geometrical effects and effects caused by the magnetic field.
Interaction effects on transient behavior have been studied in a mean-field approach
by Kurth {\it et al.}\ and Zheng {\it et al.}\ in the absence of an external
magnetic field.\cite{Kurth05:035308,Zheng07:195127}

The paper is organized as follows: Section II presents the model and the formalism, Section III
contains the numerical simulations and their extensive discussion. We conclude in Section IV.

%
\section{Model}
We consider a quantum wire with parabolic confinement potential
$V_{\mathrm{conf}}(y)=m^*\Omega_0^2y^2/2$ in a perpendicular
homogeneous magnetic field ${\mathbf B}=B\hat{\mathbf z}$.
In the Landau gauge ${\mathbf A}=-By\hat{\mathbf x}$ the time evolution of
a wave function
\begin{equation}
      \Psi ({\mathbf r},t) = \int \frac{dp}{2\pi}\frac{d\omega'}{2\pi}
      e^{i(px-\omega't)} \Psi (p,y,\omega')
\end{equation}
representing electrons impinging on the time-dependent
scattering potential $V_{\mathrm{sc}}(x,y,t)$ is determined by
the Schr{\"o}dinger equation
\begin{align}
      i\hbar\partial_t\Psi ({\mathbf r},t) = \bigg\{&-\frac{\hbar^2}{2m^*} \left(\nabla^2-
      \frac{2i}{l^2}y\partial_x -\frac{y^2}{l^4}\right)\nonumber \\
      &+ \frac{1}{2}m^*
      \Omega^2_0 y^2+ V_{\mathrm{sc}}(\textbf{r},t)\bigg\}\Psi ({\mathbf r},t).
\label{Schroedinger}
\end{align}
We need a Fourier transform in the $x$-direction to facilitate
the application of a scattering formalism to describe the transport in that direction,
as the presence of the magnetic field makes the Schr{\"o}dinger equation non-separable
in $(xy)$-coordinate space, but separable in a mixed momentum-coordinate space.\cite{Gurvitz95:7123}
The Fourier transform in time is our choice since 
we will be considering non-periodic time-dependent scattering potentials that
are only non-vanishing during a short interval of time.
In order to simplify the treatment of the motion of the electrons in the
$y$-direction (perpendicular to the transport) we expand the wave function
in terms of the eigen functions of the ideal wire, thus enabling the discussion
of the transport in terms of modes,
\begin{equation}
      \Psi (q,y,t) = \sum_n \varphi_n(q,t)\phi_n(q,y).
\label{Psi_xyt}
\end{equation}
$\phi_n(q,y)$ are the eigen functions of a harmonic oscillator shifted
by $y_0 = qa_w^2\omega_c /\Omega_w$, where $a_w=\sqrt{\hbar/(m^*\Omega_w)}$ is the characteristic length
scale replacing the magnetic length $l=\sqrt{\hbar c/(eB)}$, and
$\hbar\Omega_w=\hbar\sqrt{\omega_c^2+\Omega_0^2}$ is the characteristic energy replacing the cyclotron
energy $\hbar\omega_c=eB/(m^*c)$ and the confinement energy $\hbar\Omega_0$.

We will consider a monoenergetic incoming plane wave with energy $E$ in a definite mode $n$
\begin{equation}
      \Psi^0 ({\bf r},t) = \exp{\left[i(k_nx-\omega^0_{nk_n} t)\right]}\phi_n(k_n,y),
\label{Psi_xyt0}
\end{equation}
with the wave vector in band $n$ determined by the energy $k_na_w=\sqrt{2(E-E^0_n)\hbar\Omega_w/(\hbar\Omega_0)^2}$,
and the dispersion relation for the parabolic energy bands of the confinement
$E_{nq}=\hbar\omega^0_{nq}=E^0_{n}+(qa_w)^2(\hbar\Omega_0)^2/(2\hbar\Omega_w)$, with the band bottom
$E^0_{n}=\hbar\Omega_w(n+1/2)$ for $n=0,1,2,\cdots$. In the $(q\omega )$-plane the in-state corresponding
to Eq.\ (\ref{Psi_xyt0}) is represented by the wave function
\begin{equation}
      \varphi^0_m (q,\omega ) = (2\pi)^2\delta(q-k_n)\delta(\omega-\omega^0_{nq})\delta_{m,n},
\label{phi_qom_0}
\end{equation}
and the Schr{\"o}dinger equation (\ref{Schroedinger}) is transformed into an integral equation
\begin{align}
      \lbrace\hbar\omega- &\hbar\omega^0_{nq}\rbrace\varphi_n(q,\omega)\nonumber \\ =& 
      \sum_{n'}\int\frac{dp}{2\pi}\frac{d\nu}{2\pi} V^{\mathrm{sc}}_{nn'} (q,p,\omega-\nu)\varphi_{n'}(p,\nu) ,
\label{Schroedinger_Int}
\end{align}
with the matrix elements of the scattering potential
\begin{equation}
      V^{\mathrm{sc}}_{nn'} (q,p,\omega )=\int dy\:\phi^*_n(q,y)V_{\mathrm{sc}}(q-p,y,\omega )\phi_{n'}(p,y) .
\end{equation}
The form of the integral equation (\ref{Schroedinger_Int}) suggests an introduction of a
Green function
\begin{equation}
      \lbrace\hbar\omega-\hbar\omega^0_{nq}\rbrace G^n_0(q,\omega)= 1.
\label{Def_Green}
\end{equation}
Using the Schr{\"o}dinger equation for the wire with the embedded scatterer (\ref{Schroedinger_Int})
and the one for the ideal wire 
\begin{equation}
      \lbrace\hbar\omega-\hbar\omega^0_{nq}\rbrace\varphi^0_n(q,\omega)= 0,
\end{equation}
together with the definition for the Green function (\ref{Def_Green}) we can write a Lippmann-Schwinger
type integral equation in the $(q\omega )$-plane
\begin{widetext}
\begin{equation}
      \varphi_n(q,\omega)= \varphi^0_n(q,\omega)+ G^n_0(q,\omega) 
      \sum_{n'}\int\frac{dp}{2\pi}\frac{d\nu}{2\pi}
      V_{nn'}^{\mathrm{sc}} (q,p,\omega-\nu)\varphi_{n'}(p,\nu) .
\label{LS_phi}
\end{equation}
This equation has the advantage that it contains explicitly the asymptotic form of the wave function
for the in-state (\ref{phi_qom_0}) we are interested in and is thus a convenient stepping stone
into the scattering formalism. The presence of the magnetic field perpendicular to the
quasi one-dimensional quantum wire imposes on us the Lippmann-Schwinger equation (\ref{LS_phi})
as a coupled set of two-dimensional integral equations for the wave function $\varphi_n(q,\omega)$
in each channel or mode $n$. The wave function of the incoming state $\varphi^0_n(q,\omega)$ (\ref{phi_qom_0})
is not convenient for numerical calculations so we choose to use the T-matrix instead
\begin{equation}
      T_{nn'}(q\omega,p\nu) =
      V_{nn'}^{\mathrm{sc}}(q,p,\omega-\nu)+\sum_{m'}\int\frac{dk}{2\pi}\frac{d\omega'}
      {2\pi}V_{nm'}^{\mathrm{sc}}(q,k,\omega-\omega')G^{m'}_0(k\omega')T_{m'n'}(k\omega',p\nu),
\label{LS_Tnm_qom}
\end{equation}
from which it is easy to obtain the wave functions and make further connections
to scattering theory 
\begin{equation}
      \varphi_n(q,\omega)= \varphi^0_n(q,\omega)+G^n_0(q,\omega)\sum_{n'}\int\frac{dp}{2\pi}\frac{d\nu}{2\pi}
      T_{nn'}(q\omega,p\nu)\varphi^0_{n'}(p\nu) .
\label{LS_varphi}
\end{equation}
\end{widetext}

We are able to consider a scattering potential of the general form,
separable in time
\begin{equation}
      V_{\mathrm{sc}}({\mathbf r},t) = V({\mathbf r})F(t) ,
\label{Vxyt}
\end{equation}
in order to model the effects of a pulsed gate or a focused
microwave pulse with the time-dependent part satisfying
\begin{equation}
      F(t) = e^{-\gamma t^2} \cos(\Omega t) ,
\label{F_t}
\end{equation}
and the smooth spatial part
\begin{equation}
       V_{\mathrm{sc}}({\mathbf r})=V_0e^{-\beta r^2} .
\end{equation}
The scattering event is thus inelastic in a $2\oplus 1$-dimensional $(x,y,t)$-space,
the momentary appearance of the scattering potential (\ref{Vxyt}) can cause
a back- or forward scattering of a monoenergetic wave into a pulse with
energy spreading.

When solving the Lippmann-Schwinger equation (\ref{LS_Tnm_qom}) we use the
methods described earlier\cite{Bardarson04:01,Gudmundsson05:BT} in order to
obtain analytically the contribution of the poles of the Green function and perform
the remaining principal part integration by removing the
singularity by a subtraction of a zero.\cite{Haftel1:70,Landau:96}
The main difference from the solution of the corresponding equation
in the static case is that here in the dynamic case the evanescent
states are explicitly present in the time-dependent Green function
(\ref{Def_Green}), but in the static case they had to be included by
remembering that the $q^2$ terms there can have either sign depending on
whether they refer to a propagating state with a real wave vector or an
evanescent state with an imaginary one.

\begin{figure}[htbq]
      \includegraphics[width=0.42\textwidth,angle=0]{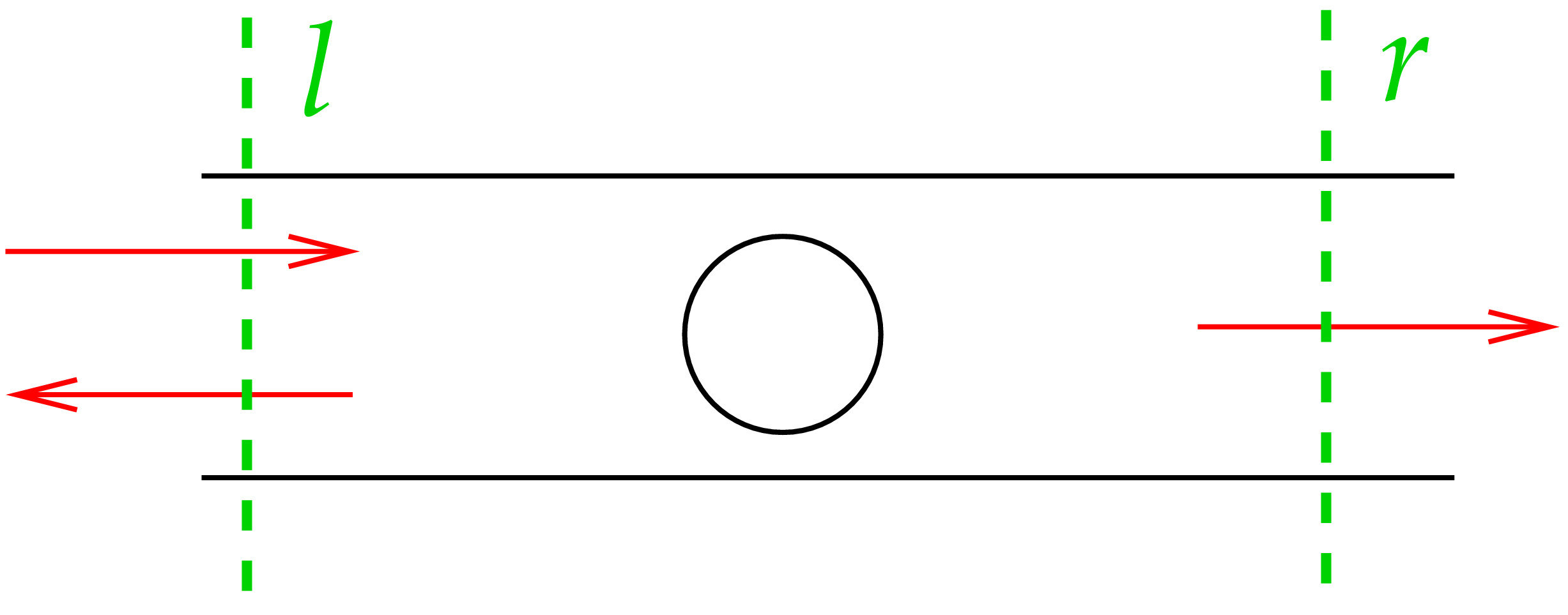}
      \caption{(Color online) A schematic figure of the system showing where the {\em left} ($l$)
               and {\em right} ($r$) currents are calculated from their densities. The in-state
               is assumed here to enter from the left.}
      \label{lr-current}
\end{figure}

We can now assemble the exact wave function together with the information
from Equations (\ref{Psi_xyt}) and (\ref{LS_varphi})
\begin{widetext}
\begin{equation}
      \Psi({\mathbf r},t)= e^{ik_nx-i\omega^0_{nk} t} \phi_n(k_n,y)+\sum_m
      \int \frac{dq}{2\pi}\frac{d\omega}{2\pi} e^{iqx-i\omega t} 
      G^m_0(q\omega)T_{mn}(q\omega,k_n\omega^0_{nk})\phi_m(q,y).
\label{Psi_xyt_i+sc}
\end{equation}
\end{widetext}
We have the option to continue along two different paths; The wave function
(\ref{Psi_xyt_i+sc}) is composed of the incoming wave and the scattered one.
Traditionally in a scattering calculation one would seek the asymptotic limit
by first Fourier transforming the T-matrix
\begin{equation}
      T_{mn}(qt,pt') = \int\frac{d\omega}{2\pi}\frac{d\nu}{2\pi} e^{-i\omega t} T_{nm}(q\omega,p\nu) e^{i\nu t'},
\end{equation}
or rather Fourier transforming (to enhance the convergence)
\begin{equation}
      T_{nm}(q\omega ,p\nu )-V_{nm}(q,p)f(\omega ,\nu ),
\end{equation}
with $f(\omega ,\nu )$ being the double Fourier transform of $f(t,t')=F(t)\delta (t-t')$,
in order to obtain the transition probability 
\begin{widetext}
\begin{equation}
      t_{mn}(t) = \delta_{mn}-\frac{1}{a^2_w}
      \left(\frac{\hbar\Omega_w}{\hbar^2\Omega^2_0} \right)\frac{i}{2k_m}\int^t_{-\infty} dt'
      e^{-\omega^0_{mk_m}(t'-t)}T_{mn}(k_m t,k_n t') .
\end{equation}
\end{widetext}
Since we are neither modeling here a static nor periodic system there is no immediate connection
of $t_{mn}(t)$ to conductance of the system through the Landauer-B{\"u}ttiker approach, but we could
use it to calculate the asymptotic wave functions needed to calculate the currents in and out
of the system. Within the accuracy of the numerical approach it is also possible to skip these
steps and use directly the wave function (\ref{Psi_xyt_i+sc}) evaluated in the asymptotic
regions of the quantum wire to define the current. This is schematically presented in 
Fig.\ \ref{lr-current} for an in-state entering the system from the left.
In the Landau gauge the {\em left} and {\em right}
particle currents for state $\alpha = |nq\rangle$ (calculated in the asymptotic region where
$V_{\mathrm sc}$ (\ref{Vxyt}) is vanishing) of interest are
\begin{equation}
      (I_\alpha^{r,l}(t))_x = \frac{\hbar}{m^*}\Re\left\{\int_{-\infty}^{\infty}
                     dy (\Psi_\alpha^{r,l})^* D_x\Psi_\alpha^{r,l}\right\} ,
\label{I_x}
\end{equation}
with $\hbar D_x=(p_x+(e/c)A_x)=\hbar (-i\partial_x-y/l^2)$.
In an unbiased ideal quantum wire with no scattering center 
the left- and right-going currents will cancel
each other. Here we consider the left asymptotic region of the wire as the
source electrode with its chemical potential $\mu_l=\mu+\delta\mu$ raised
by an infinitely small amount $\delta\mu$ above the chemical potential
in the right asymptotic region $\mu_r=\mu$. For the scattering potential $V_{\mathrm sc}$
considered here with full $(x,y,t)$-reflection symmetry it is thus sufficient to calculate the
current for the in-state with energy $E=\mu_l$. The states with $E\leq\mu_r$ 
do not contribute to the {\em net} current through the system, but the time variation of
the spatial and temporal symmetric scattering potential (\ref{Vxyt}) induces
currents in opposite directions to the left and right of the scattering center.
Before the appearance of the scattering center these currents are nonexistent, but
they can remain after its disappearance. We shall not be concerned with these currents
since, as stated before, they do not contribute to the net current through the
wire induced by $\delta\mu$ and the variation of $V_{\mathrm sc}$. 

In the static case the conductance is calculated through the Landauer-B{\"u}ttiker
formalism using the probability amplitude $t_{nm}(E)$ for the transition from an
in-state $|mk_m(E)\rangle$ to any available propagating outstate $|nk_n(E)\rangle$
with the energy $E$ conserved. Of course all off-shell intermediate states are
present in the T-matrix reflecting the multiple scattering character of the
Lippmann-Schwinger formalism. The effects of all these states, propagating or not
are found in the exact wave function (\ref{Psi_xyt_i+sc}) both in the static and the
dynamical case. Here, in the dynamic case the scattering is nonelastic meaning that
an incoming electron with definite energy can leave the system in a state with
different energy. Our incoming electron is not localized in space, it has a definite
energy $E$ and is represented by a wave function with a plane wave component.
The scattering process spreads the energy of this incoming state and thus allows
for the formation of a wave packet or a pulse that then propagates in the wire.

We use the eigen functions $\phi_{n}(q,y)$ as a nonorthogonal basis and
a grid in time $t$ and the Fourier variable $q$
to cast the integral equation for the T-matrix $T_{nn'}(q\omega,p\nu)$ (\ref{LS_Tnm_qom})
into a set of linear equations to be solved numerically.
The grids are constructed to enable the implied integrations to be performed by a
repeated 4-point Gaussian method for numerical integration.

\section{Results and discussion}
\begin{figure}[htbq]
      \includegraphics*[width=0.42\textwidth,angle=0,viewport=10 10 345 198]{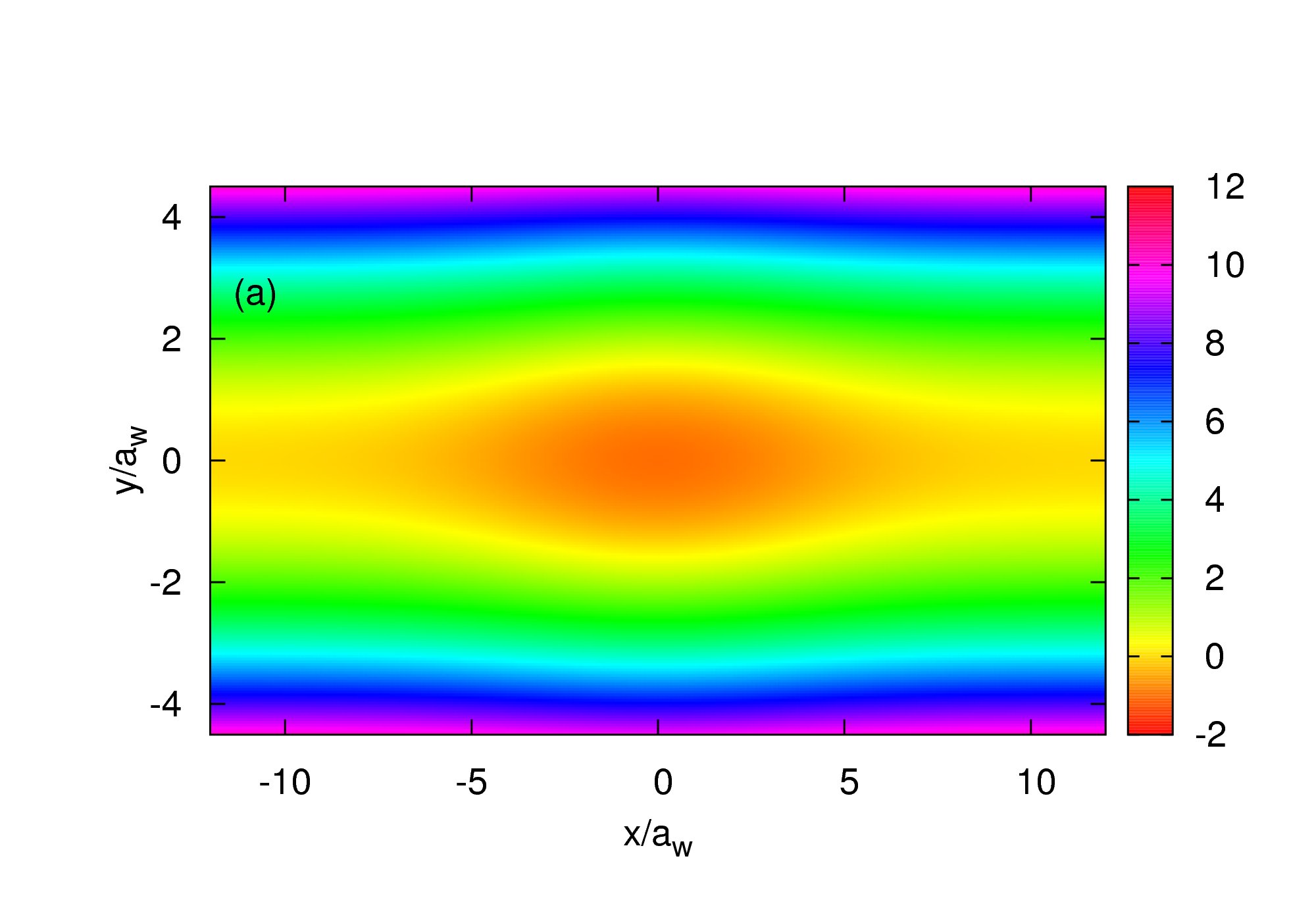}
      \includegraphics*[width=0.42\textwidth,angle=0,viewport=10 18 345 198]{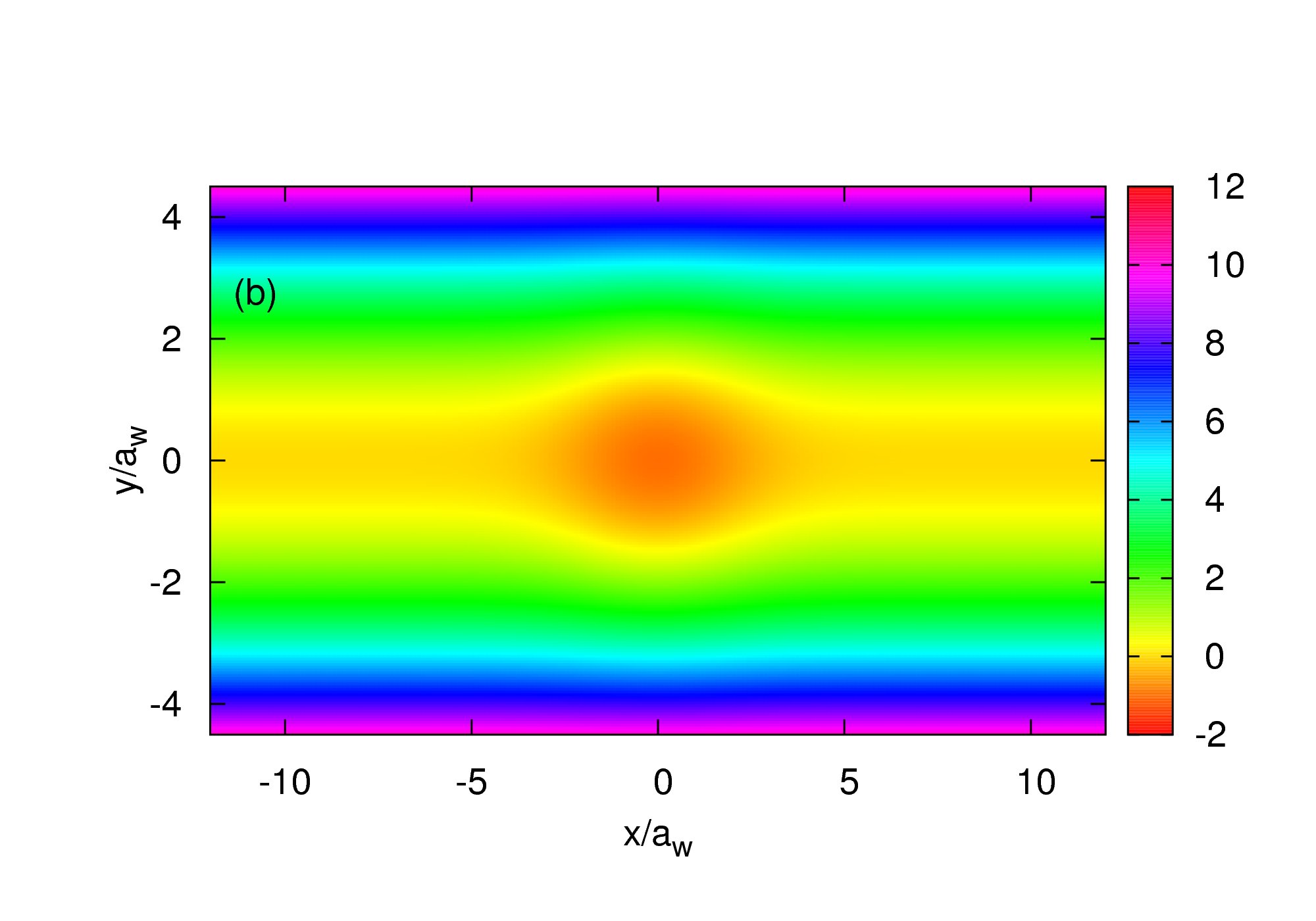}
      \caption{(Color online) The scattering potential $V_{\mathrm{sc}}({\mathbf r},0)$ at
                              $t=0$ for 
                              (a) $\beta =1\times 10^{-4}$ nm$^{-2}$,
                              (b) $\beta =4\times 10^{-4}$ nm$^{-2}$; $V_0=-1.0$ meV.}
      \label{Vsc_D}
\end{figure}

\begin{figure}[htbq]
      \includegraphics*[width=0.42\textwidth,angle=0,viewport=10 10 345 198]{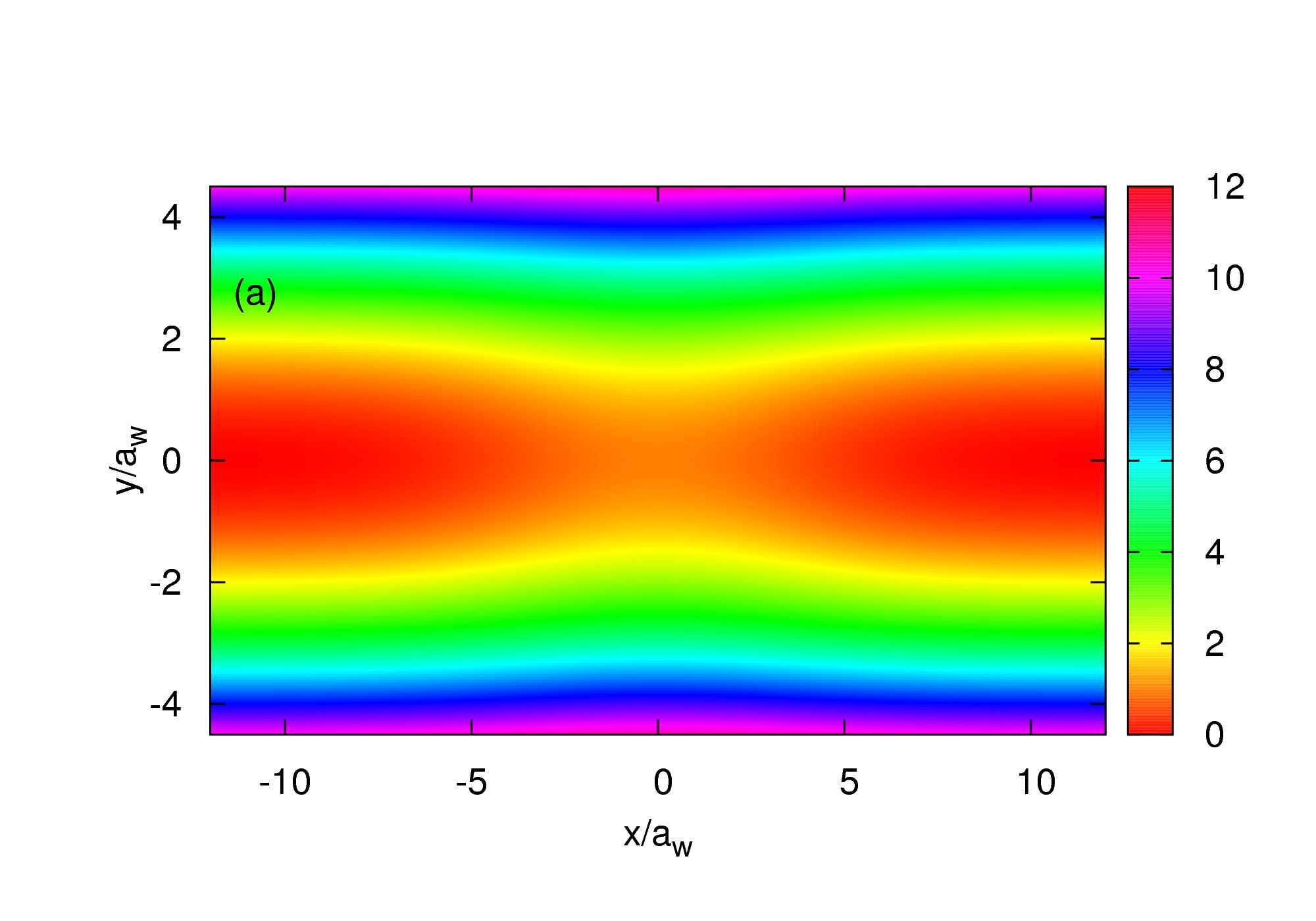}
      \includegraphics*[width=0.42\textwidth,angle=0,viewport=10 18 345 198]{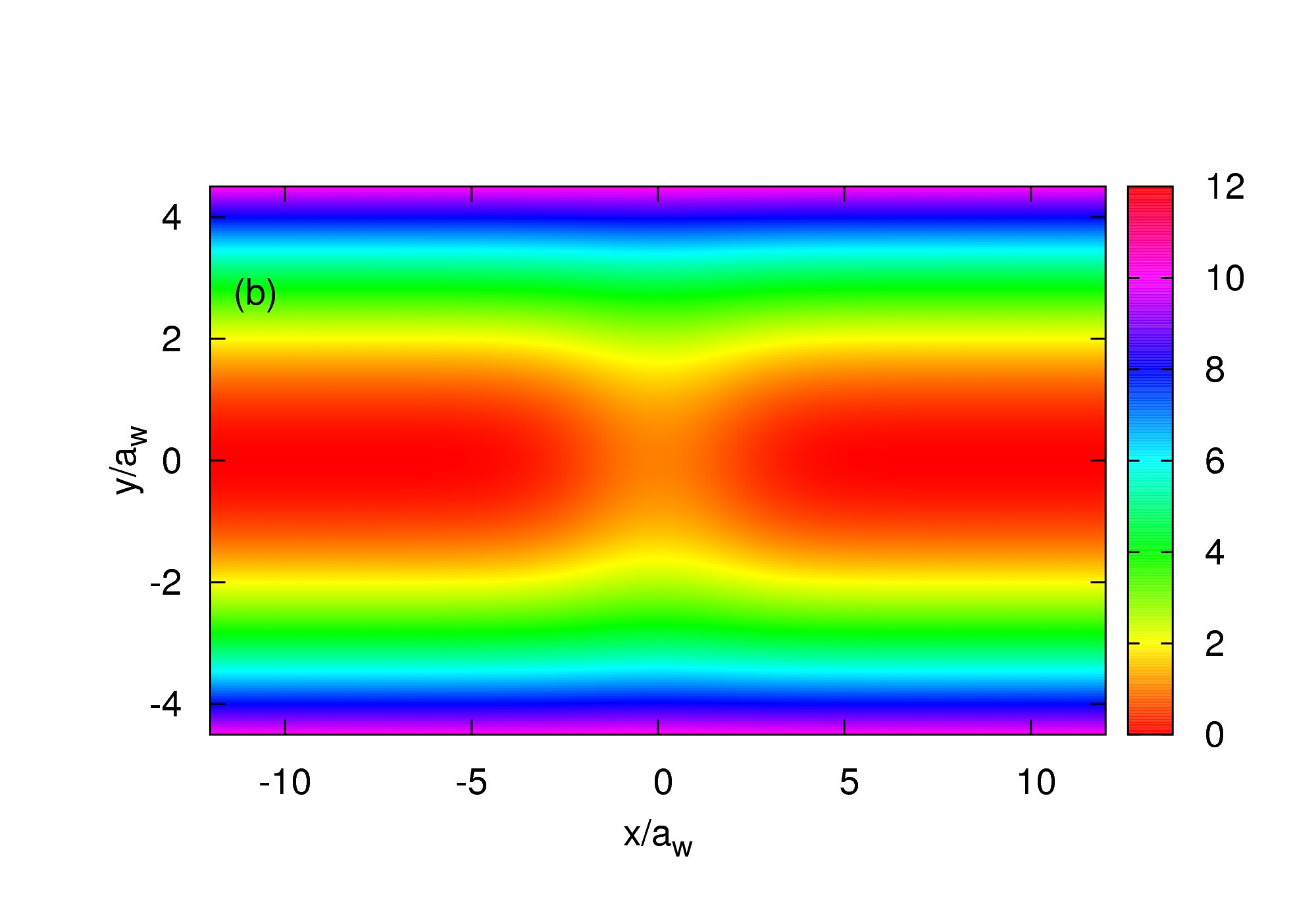}
      \caption{(Color online) The scattering potential $V_{\mathrm{sc}}({\mathbf r},0)$ at
                              $t=0$ for 
                              (a) $\beta =1\times 10^{-4}$ nm$^{-2}$,
                              (b) $\beta =4\times 10^{-4}$ nm$^{-2}$; $V_0=+1.0$ meV.}
      \label{Vsc_H}
\end{figure}

We will consider the spatial part of the scattering potential $V_{\mathrm{sc}}$ to 
be a smooth Gaussian well or hill (\ref{Vxyt}) with a choice of strength
($V_0=\pm 1.0$ meV) and width ($\beta =1\times 10^{-4}$ or $4\times 10^{-4}$ nm$^{-2}$)
to form a temporary shallow well seen in Fig.\ \ref{Vsc_D}, or a temporary quantum 
constriction displayed in Fig.\ \ref{Vsc_H}. 

\begin{figure}[htbq]
      \includegraphics*[width=0.42\textwidth,angle=0,viewport=0 10 360 245]{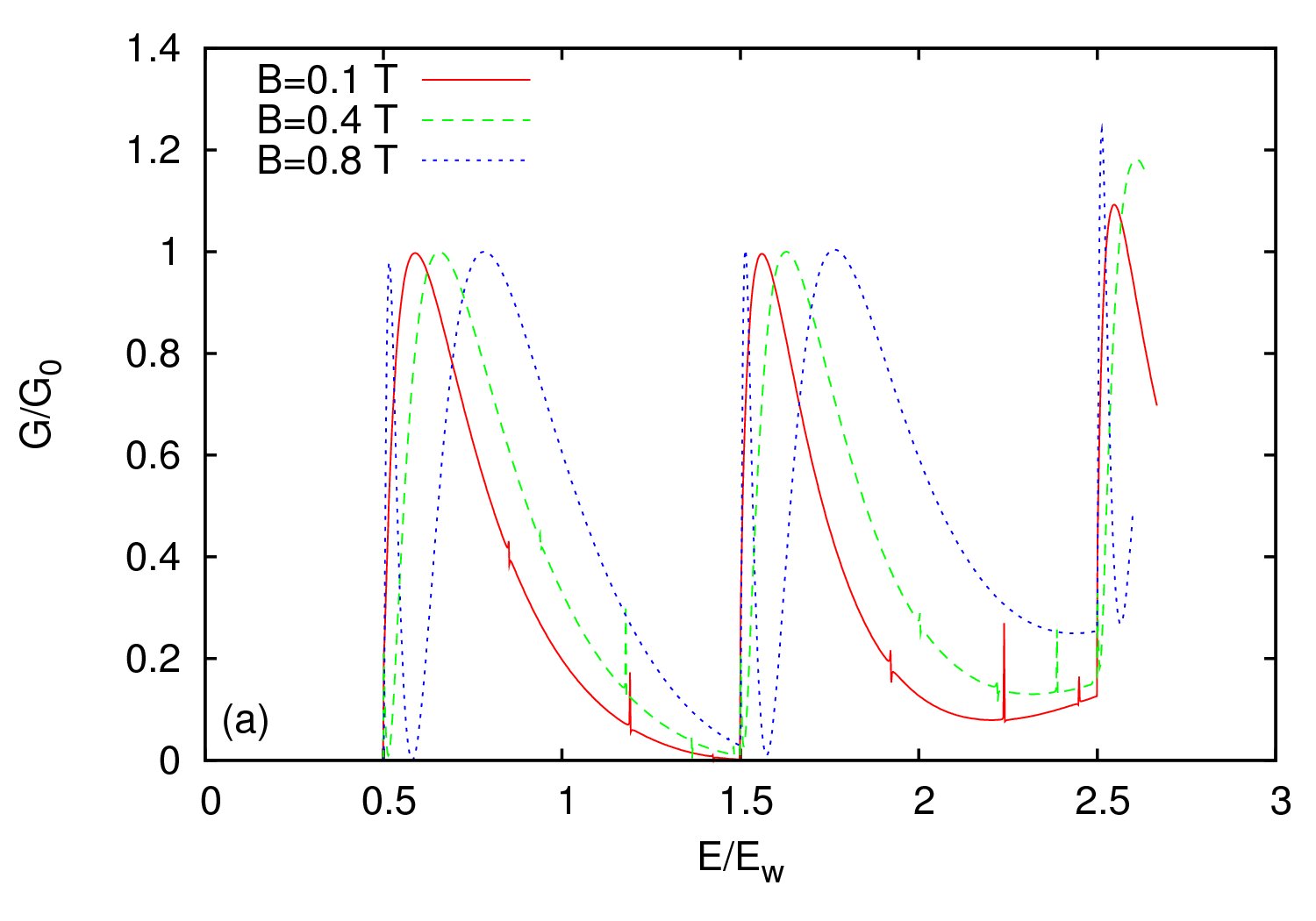}
      \includegraphics*[width=0.42\textwidth,angle=0,viewport=0 10 360 245]{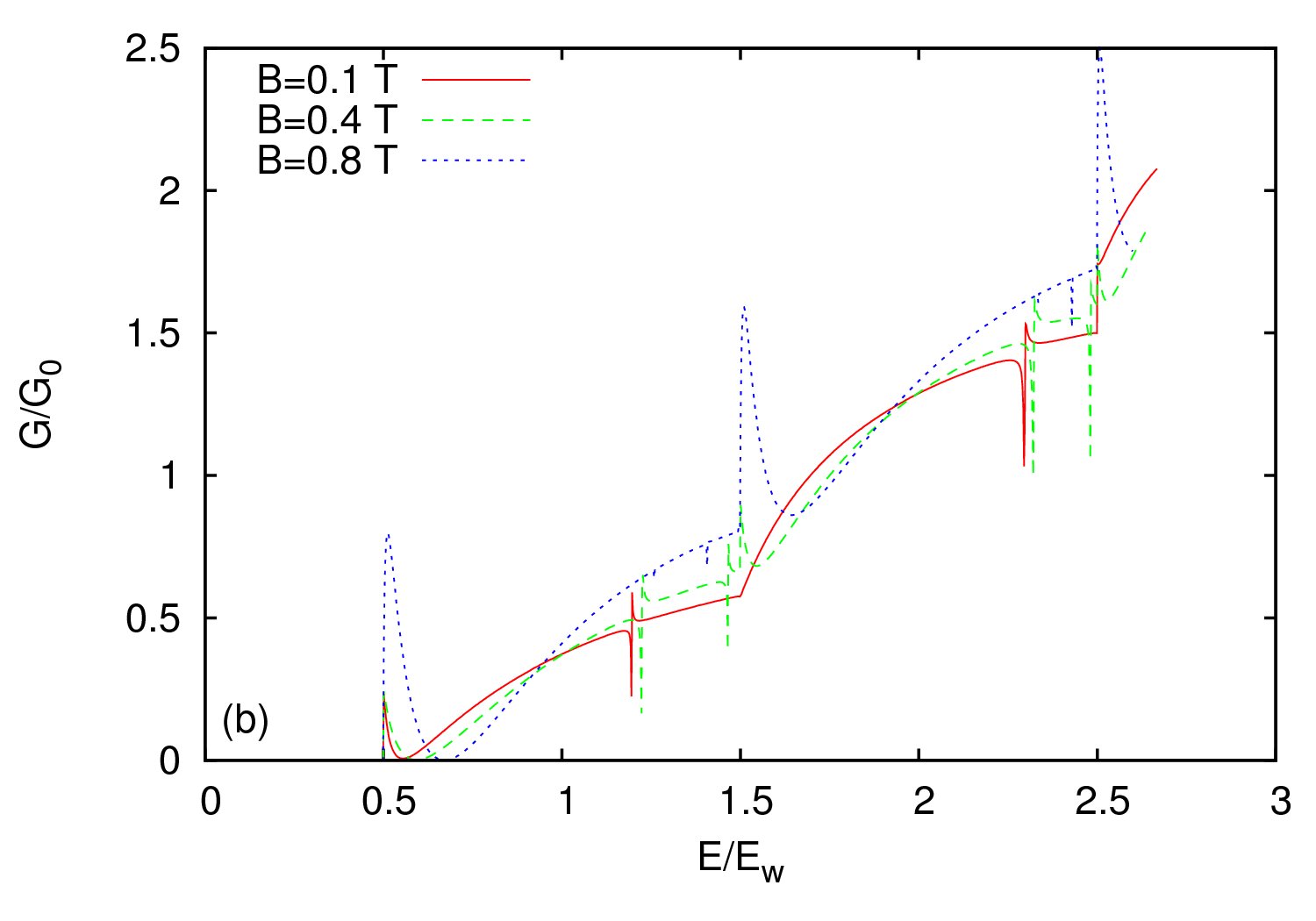}
      \caption{(Color online) The conductance of the static system vs.\ $E/E_w=E/(\hbar\Omega_w)$
               with scattering potential $V_{\mathrm{sc}}({\mathbf r},0)$ for
               (a) $\beta =1\times 10^{-4}$ nm$^{-2}$,
               (b) $\beta =4\times 10^{-4}$ nm$^{-2}$, $V_0=-1.0$ meV.}
      \label{GE_D}
\end{figure}

\begin{figure}[htbq]
      \includegraphics*[width=0.42\textwidth,angle=0,viewport=0 10 360 245]{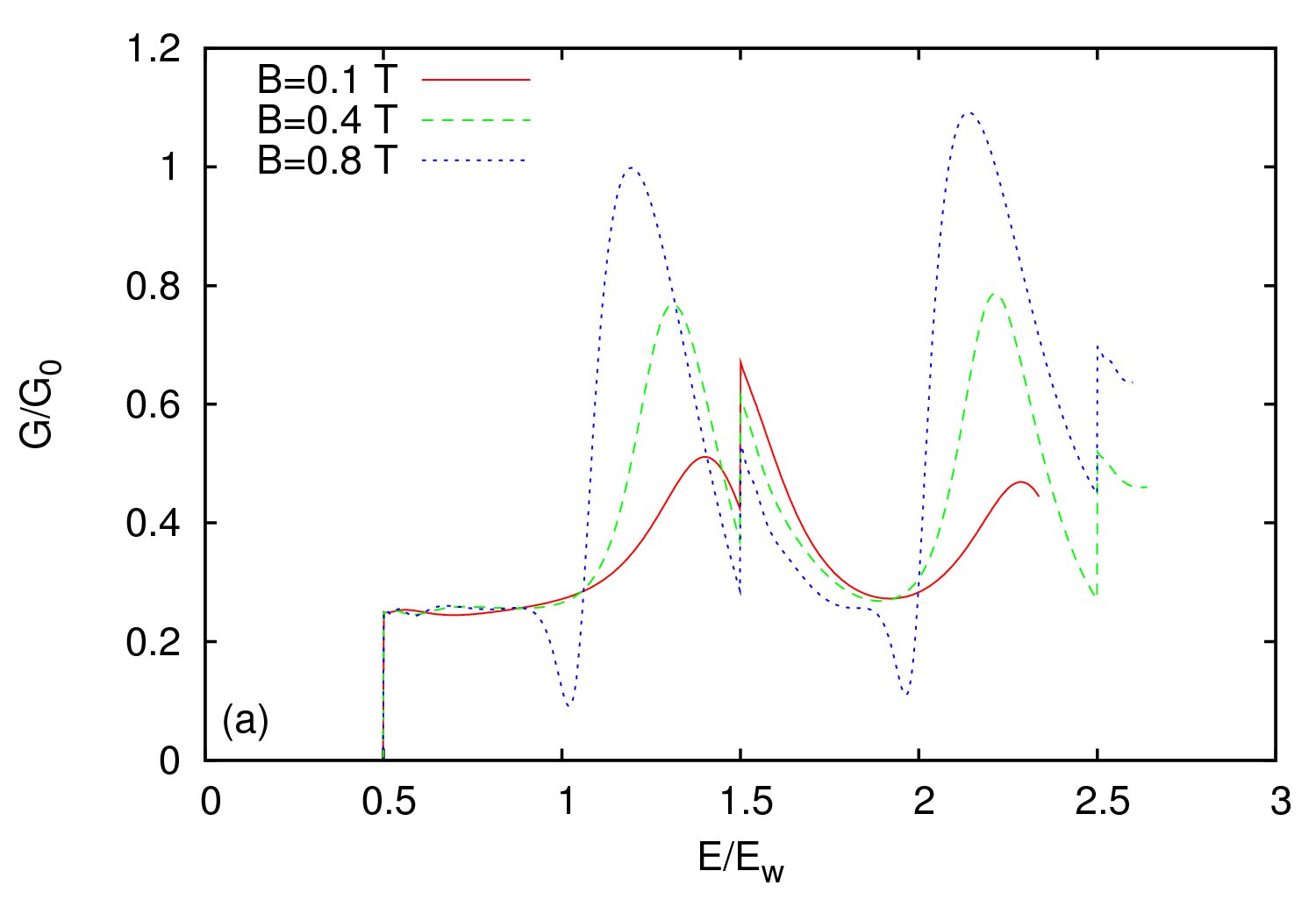}
      \includegraphics*[width=0.42\textwidth,angle=0,viewport=0 10 360 245]{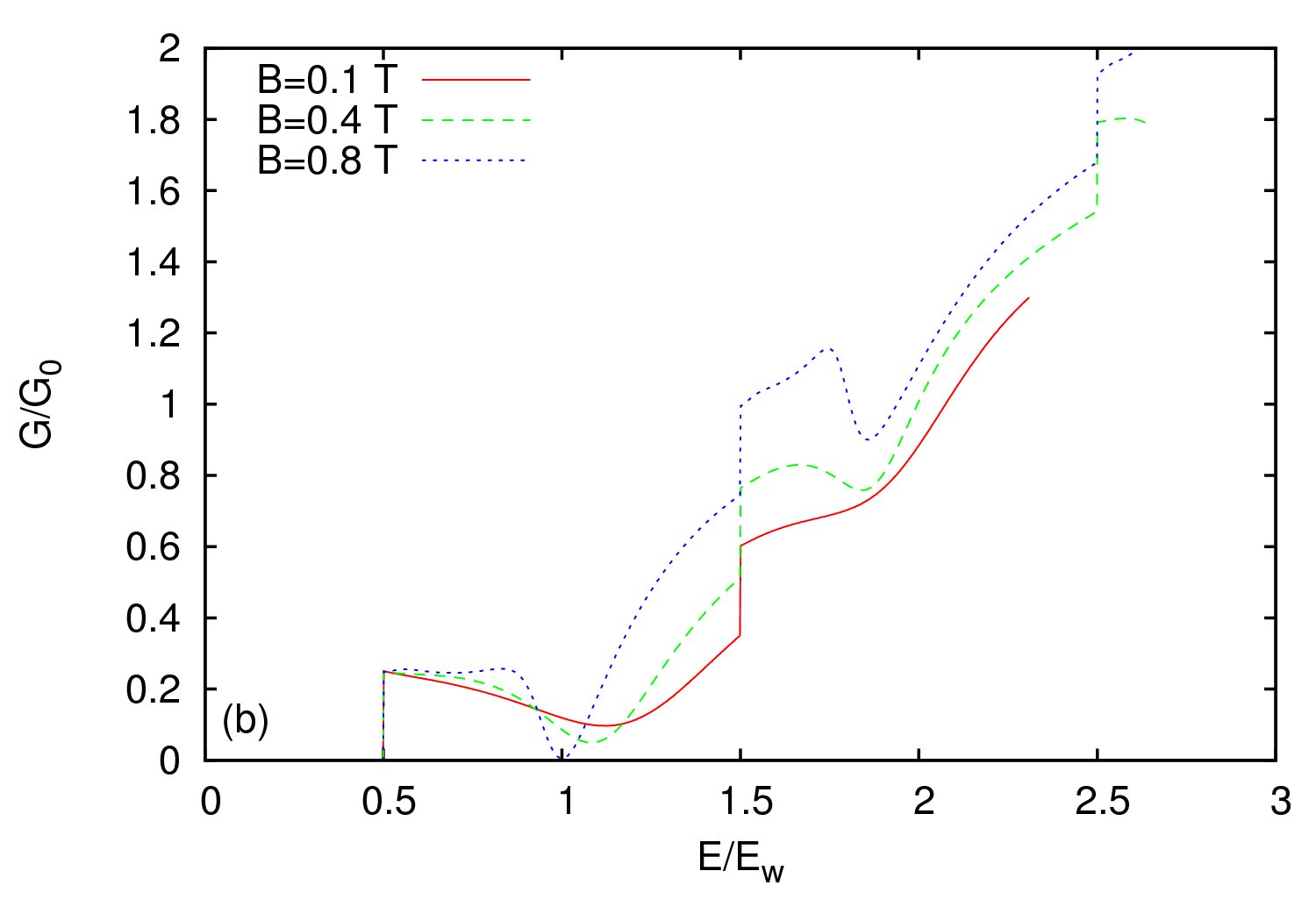}
      \caption{(Color online) The conductance of the static system vs.\ $E/E_w=E/(\hbar\Omega_w)$
               with scattering potential $V_{\mathrm{sc}}({\mathbf r},0)$ for
               (a) $\beta =1\times 10^{-4}$ nm$^{-2}$,
               (b) $\beta =4\times 10^{-4}$ nm$^{-2}$, $V_0=+1.0$ meV.}
      \label{GE_H}
\end{figure}

In Figures \ref{GE_D} and \ref{GE_H} we show the static conductance\cite{Gudmundsson05:BT}
of the systems we will be studying the dynamical behavior of, the Gaussian potential
well and hill, respectively. The characteristic length of the narrower well seen
in Fig.\ \ref{GE_D}(b) is small enough not to show any signs of Aharanov-Bohm
oscillations for the regime of low magnetic field strength we are considering.
On the other hand the onset of the Aharanov-Bohm oscillations for the broader well
in Fig.\ \ref{GE_D}(a) with the ``wavelength'' exceeding the width of the
conductance plateau for the lowest values of the magnetic field gives the
conductance curve for the broader well a totally different character.
The characteristic steps are absent and will only return for still
higher energies. In addition to the washed out steps and the Aharanov-Bohm oscillations
we also see very narrow Fano-like resonances caused by quasi-bound states of the
wells occurring in the continuous energy spectrum of the quantum wire.
\cite{Gudmundsson05:BT} 

The potential landscape of the quantum constrictions (Fig.\ \ref{Vsc_H}) does not include
any closed disconnected contours around ``hills'' allowing for quasi-bound states
with negative energy \cite{Gudmundsson05:153306} and thus no sharp resonance features are
seen in the conductance through them shown in Fig.\ \ref{GE_H}. Instead broad features
are present caused by resonances with short life-time.

For the time-dependent potential (\ref{F_t}) we select $\Omega =0.2\Omega_w$, and
$\gamma = 1.0\Omega_w^{-2}$, leading to a smooth flashing of the scattering potential
in a Gaussian manner with the pulse reaching more than half width only inside the
interval $t\Omega_w\in [-1,+1]$.

\begin{figure}[htbq]
      \includegraphics*[width=0.42\textwidth,angle=0,viewport=4 20 345 198]{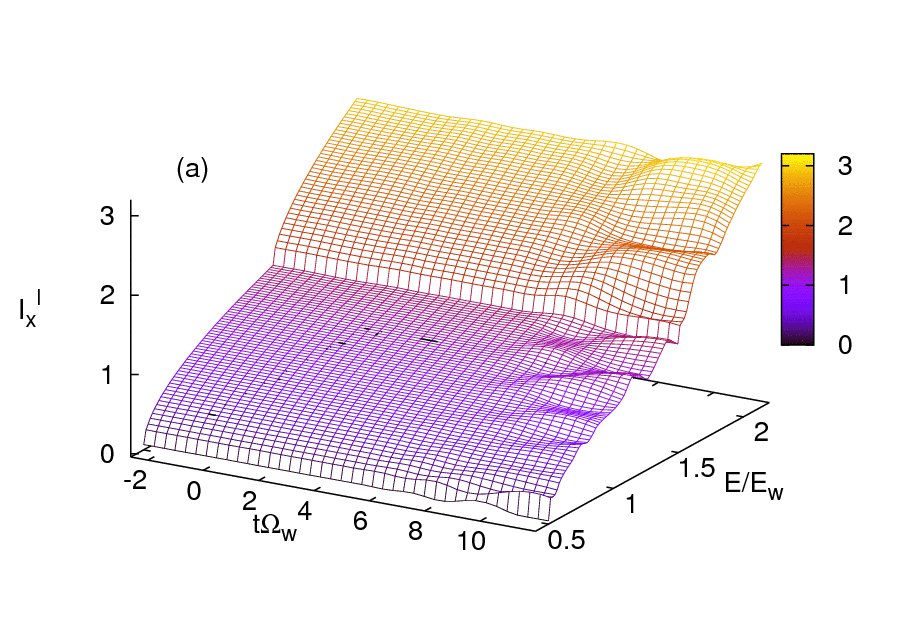}
      \includegraphics*[width=0.42\textwidth,angle=0,viewport=4 20 345 198]{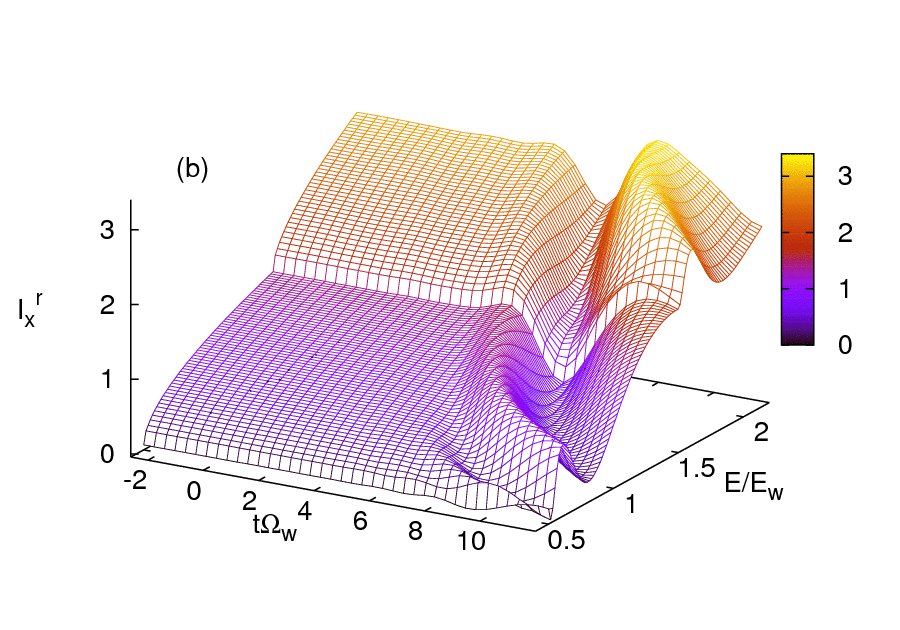}
      \caption{(Color online) The current $I_x^l(t)$ (a), and $I_x^r(t)$ (b)
               for $B=0.1$ T, $\beta =4\times 10^{-4}$ nm$^{-2}$, $V_0=-1.0$ meV,
               $\Omega =0.2\Omega_w$, and $\gamma = 1.0\Omega_w^{-2}$.}
      \label{JB01_b04_Vm1}
\end{figure}

The effects of the time-dependent potential (\ref{Vxyt}) on the transport through the
wire can be best observed by monitoring the {\em net} time-dependent current 
as a function of time $t$ and energy $E$ of the in-state. This is done in Fig.\ \ref{JB01_b04_Vm1}
for a Gaussian well flashed on and off at $B=0.1$ T. 
Both the left $I_x^l(t)$ and the right current $I_x^r(t)$ show the characteristic
square root increase with $E$ and the onset of a second subband typical for
an ideal parabolically confined quantum wire within the energy range shown here.
In an addition, $I_x^l(t)$ shows well after the disappearance of the scattering
potential a weak interference pattern emanating from the scattering center to the
left. This is caused by a small backscattered wave interfering with the incoming
wave and the variation with $E$ reflects the change in the wave vectors of the
two waves. A simplistic comparison with the static conductance in Fig.\ \ref{GE_D} 
might leave one to wonder why there is not stronger backscattering at low energy.
Before answering to that concern we should look at the current to the right
of the scattering center $I_x^r(t)$. It shows strong modulation propagating 
along the wire to the right. The scattering potential is short lived here with
a maximum strength at $t=0$. It can strongly modulate the net current in the
wire by delaying the transport of, or by pushing the probability temporarily 
around the scattering center, but it does not cause a strong permanent backscattering. 

\begin{figure}[htbq]
      \includegraphics*[width=0.42\textwidth,angle=0,viewport=4 20 345 198]{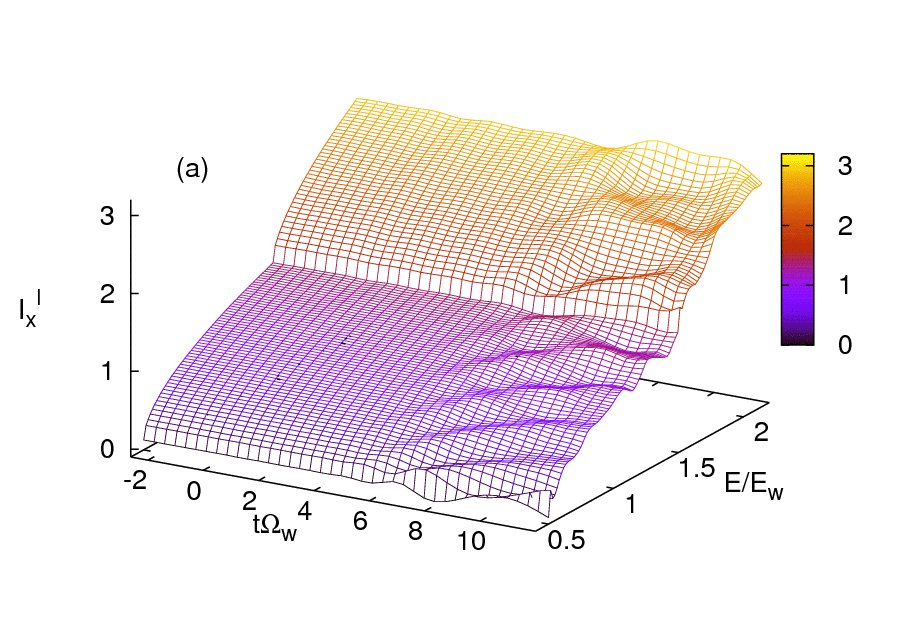}
      \includegraphics*[width=0.42\textwidth,angle=0,viewport=4 20 345 198]{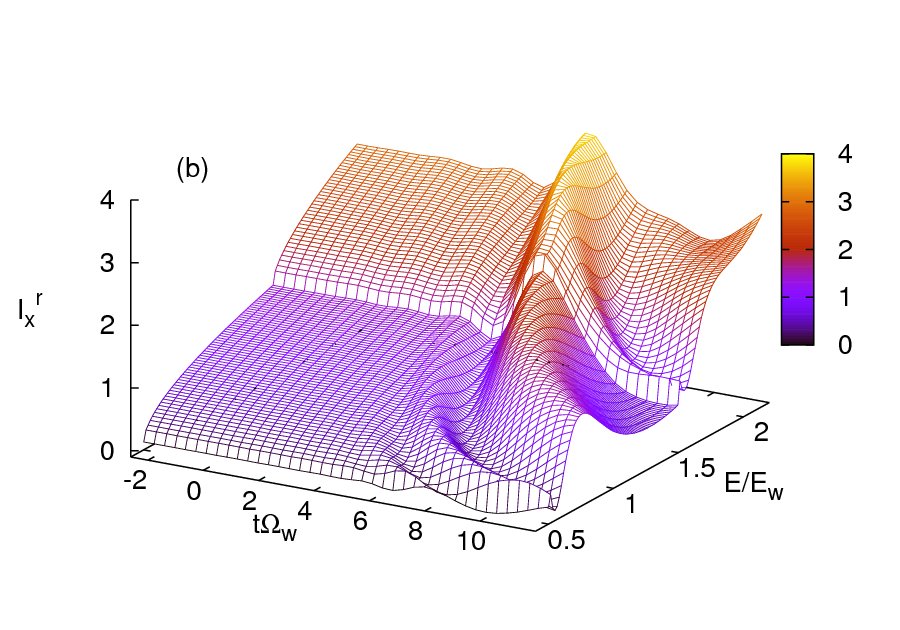}
      \caption{(Color online) The current $I_x^l(t)$ (a), and $I_x^r(t)$ (b)
               for $B=0.1$ T, $\beta =4\times 10^{-4}$ nm$^{-2}$, $V_0=+1.0$ meV,
               $\Omega =0.2\Omega_w$, and $\gamma = 1.0\Omega_w^{-2}$.}
      \label{JB01_b04_Vp1}
\end{figure}

We started the discussion about the effects on the current by looking at the
effects of the flashing on and off by a very smooth Gaussian well potential
seen in Fig.\ \ref{Vsc_D}(a). The effects on the current by the complimentary
smooth hill potential seen in Fig.\ \ref{Vsc_H}(a) flashed on and off 
in the same manner is seen in Fig.\ \ref{JB01_b04_Vp1}.
Here a second look at the smooth hill potential seen in Fig.\ \ref{Vsc_H}(a)
convinces us that actually the smooth hill with the parameters here represents
a quantum constriction, and the parameter $\beta =4\times 10^{-4}$ nm$^{-2}$ that
earlier lead to a smooth well now causes the constriction to be more extended than
the lower value $\beta =1\times 10^{-4}$ nm$^{-2}$  Indeed, we notice immediately
that the effects on the current by the long constriction are stronger than 
the effects of the smooth well if we judge only from the stronger interference
seen between the backscattered wave and the incident wave in $I_x^l(t)$ in
Fig.\ \ref{JB01_b04_Vp1}(a).

\begin{figure}[htbq]
      \includegraphics*[width=0.42\textwidth,angle=0,viewport=4 20 355 198]{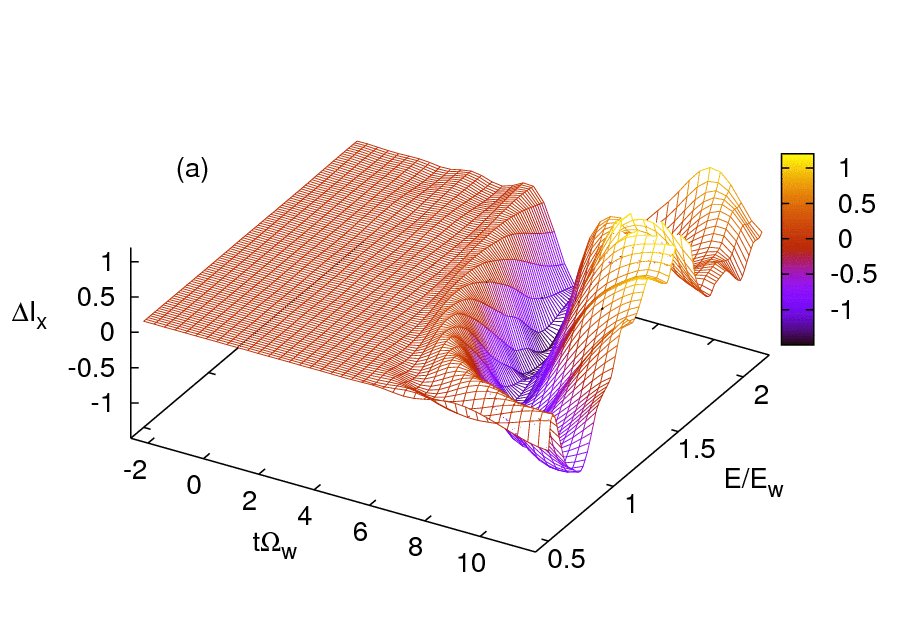}
      \includegraphics*[width=0.42\textwidth,angle=0,viewport=4 20 355 198]{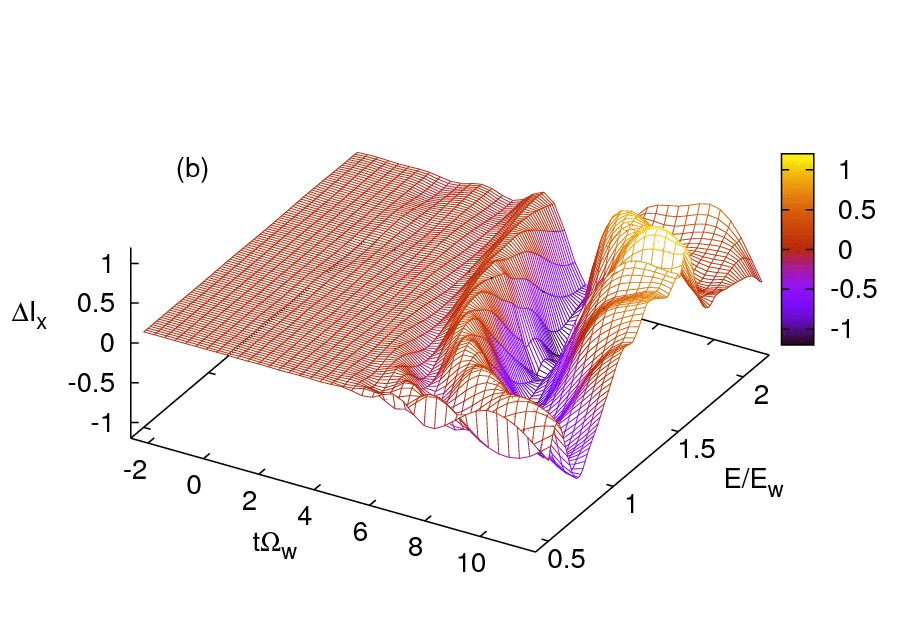}
      \caption{(Color online) The difference of net currents $\Delta I_x(t)=I_x^l(t)-I_x^r(t)$
               into the scattering region of the system for
               $\beta =1\times 10^{-4}$ nm$^{-2}$ (a),
               $\beta =4\times 10^{-4}$ nm$^{-2}$ (b).
               $B=0.1$ T, $V_0=+1.0$ meV,
               $\Omega =0.2\Omega_w$, and $\gamma = 1.0\Omega_w^{-2}$.}
      \label{DJB01p}
\end{figure}

In order to compare better the influence on the net current by the various potentials 
flashed on and off in the wire we turn our attention to the difference of net currents momentarily 
entering the scattering region in the wire $\Delta I_x(t)=I_x^l(t)-I_x^r(t)$.
\begin{figure}[htbq]
      \includegraphics*[width=0.42\textwidth,angle=0,viewport=4 20 355 198]{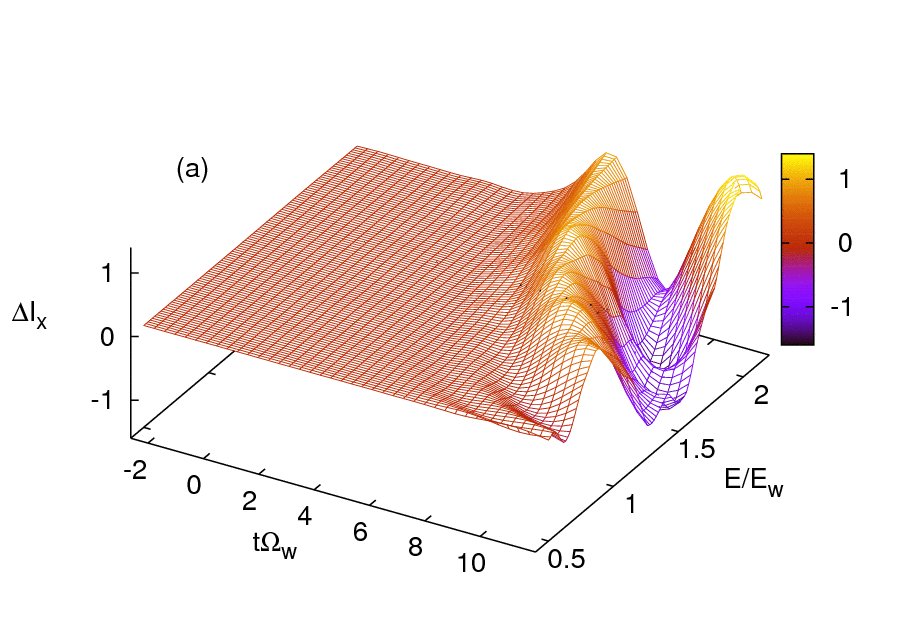}
      \includegraphics*[width=0.42\textwidth,angle=0,viewport=4 20 355 198]{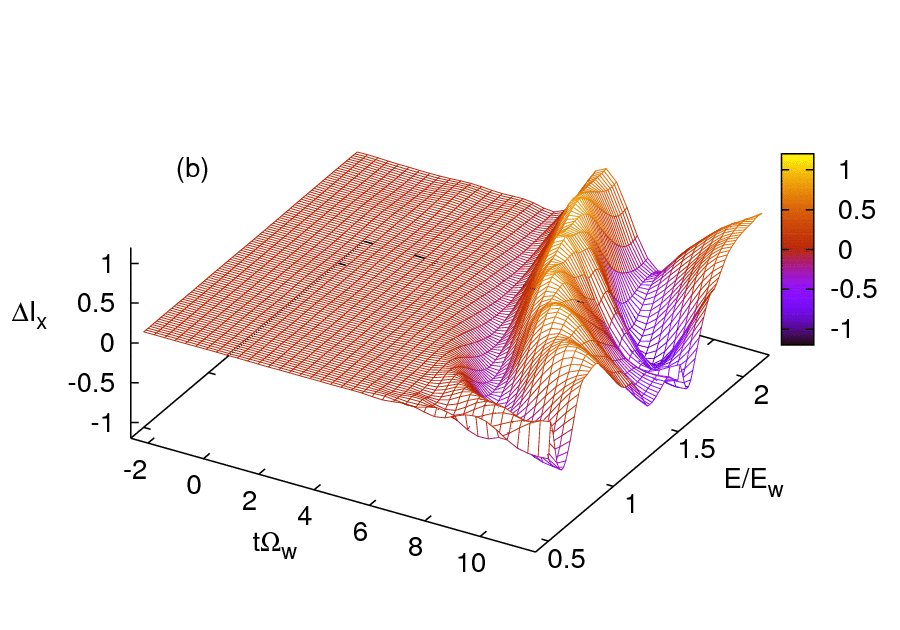}
      \caption{(Color online) The difference of net currents $\Delta I_x(t)=I_x^l(t)-I_x^r(t)$
               into the scattering region of the system for
               $\beta =1\times 10^{-4}$ nm$^{-2}$ (a),
               $\beta =4\times 10^{-4}$ nm$^{-2}$ (b).
               $B=0.1$ T, $V_0=-1.0$ meV,
               $\Omega =0.2\Omega_w$, and $\gamma = 1.0\Omega_w^{-2}$.}
      \label{DJB01m}
\end{figure}
In Fig.\ \ref{DJB01m} this is shown for both types of wells flashed on and off,
and in Fig.\ \ref{DJB01p} for the constrictions. Displaying here the difference between
the left and the right net current washes out the smooth step indicating 
the onset of transport in the second subband. Since the formation of a constriction
is dynamical it is no surprise that their effects are seen earlier in the net
current $\Delta I_x(t)=I_x^l(t)-I_x^r(t)$ as their formation causes probability
density to be ``squeezed'' out in both directions, in contrast to the formation of a
well where at least a part of the probability is ``sucked'' into the scattering
region. 

Now, one might be concerned that we actually intent showing that the
scattering region in our quantum wire can be depleted or charged momentarily
by the time-dependent scattering potential, but we have to remember that our
Lippmann-Schwinger approach here is essentially a single-electron picture.
The more correct description to use here would be to say that the pulsed potential
can either accumulate or deplete electron probability, or as we will see later,
form quasi-bound states momentarily. All reference to a many-electron picture reminds us
of the important role played by the Coulomb interaction in such dynamical
effects. 

\begin{figure}[htbq]
      \includegraphics*[width=0.42\textwidth,angle=0,viewport=4 20 355 198]{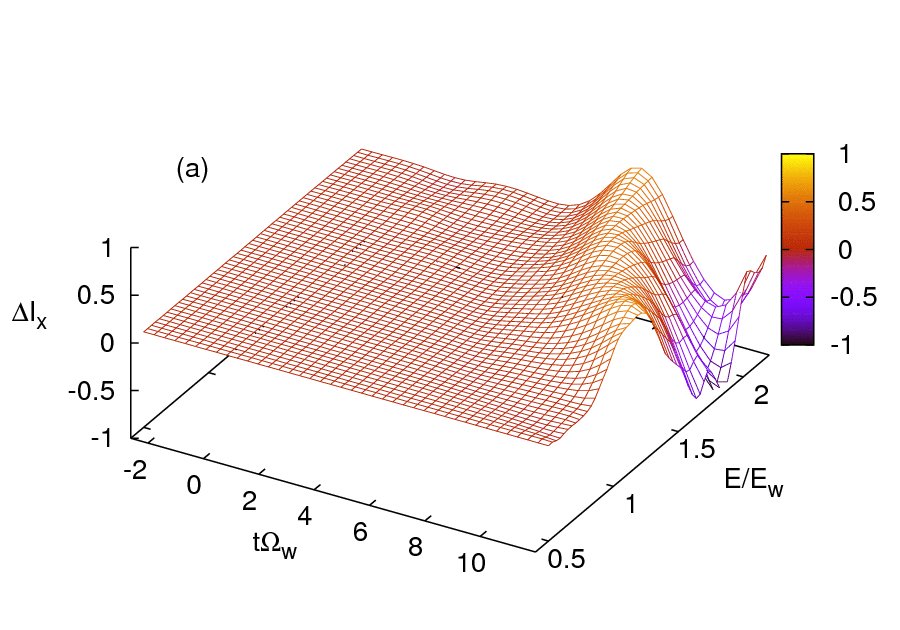}
      \includegraphics*[width=0.42\textwidth,angle=0,viewport=4 20 355 198]{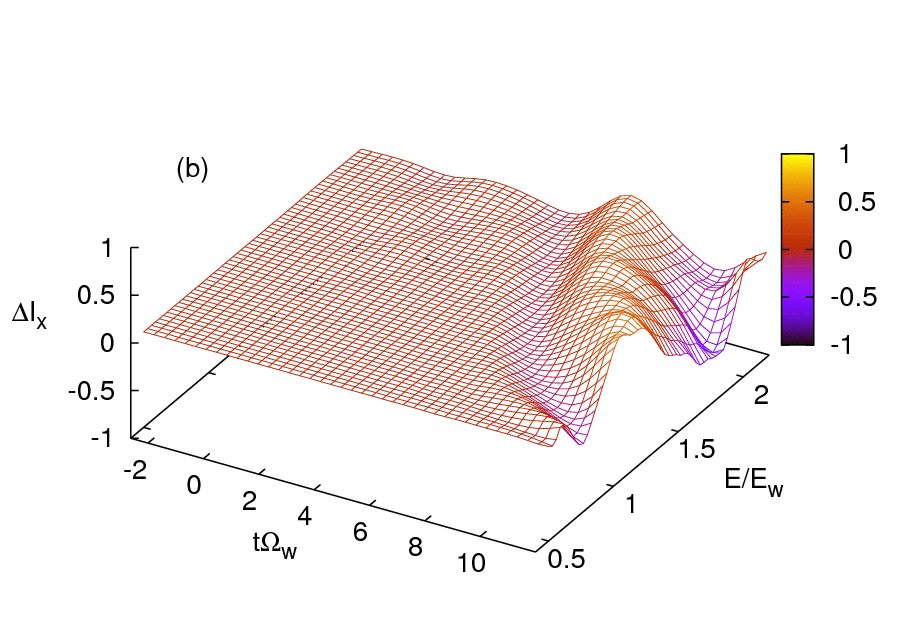}
      \caption{(Color online) The difference in net currents $\Delta I_x(t)=I_x^l(t)-I_x^r(t)$
               into the scattering region of the system for
               $\beta =1\times 10^{-4}$ nm$^{-2}$ (a),
               $\beta =4\times 10^{-4}$ nm$^{-2}$ (b).
               $B=0.4$ T, $V_0=-1.0$ meV,
               $\Omega =0.2\Omega_w$, and $\gamma = 1.0\Omega_w^{-2}$.}
      \label{DJB04m}
\end{figure}

\begin{figure}[htbq]
      \includegraphics*[width=0.42\textwidth,angle=0,viewport=4 20 355 198]{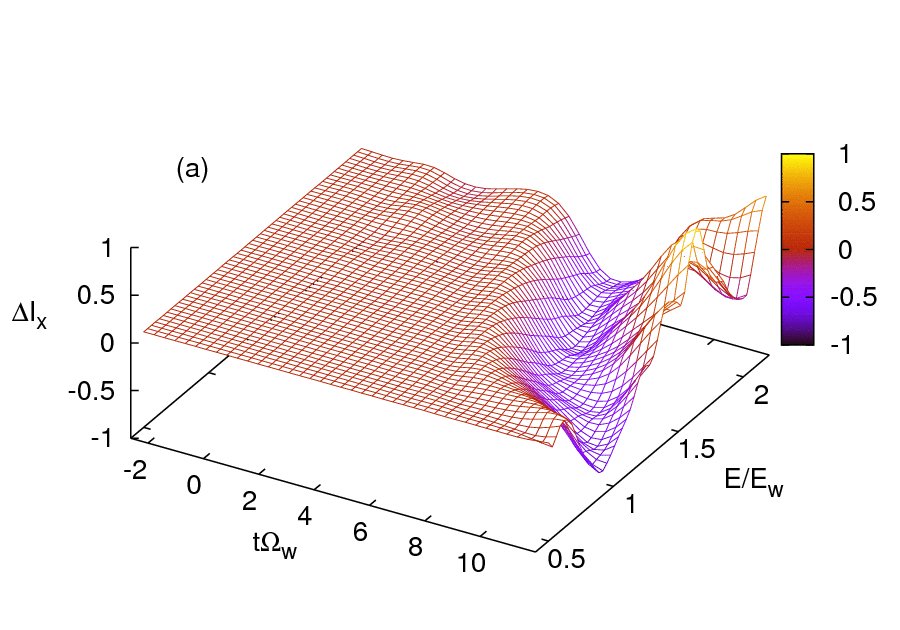}
      \includegraphics*[width=0.42\textwidth,angle=0,viewport=4 20 355 198]{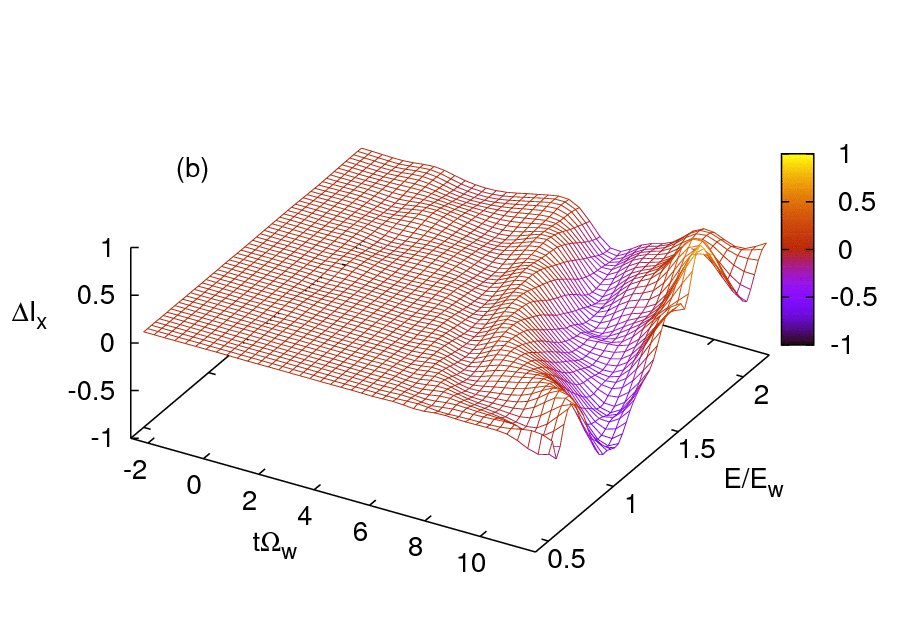}
      \caption{(Color online) The difference of net currents $\Delta I_x(t)=I_x^l(t)-I_x^r(t)$
               into the scattering region of the system for
               $\beta =1\times 10^{-4}$ nm$^{-2}$ (a),
               $\beta =4\times 10^{-4}$ nm$^{-2}$ (b).
               $B=0.4$ T, $V_0=+1.0$ meV,
               $\Omega =0.2\Omega_w$, and $\gamma = 1.0\Omega_w^{-2}$.}
      \label{DJB04p}
\end{figure}

In Fig's \ref{DJB04m} and \ref{DJB04p} we show again the net current into
the scattering center for the flashing on and off of a well and a constriction,
respectively, but now for the higher magnetic field of $B=0.4$ T.
Clearly, the effects on the current are weaker at this higher magnetic field.
The main reason behind this is the Lorentz force shifting the incoming plane
wave (in the lowest wire mode) away from the middle of the wire where the
potential will have its maximum at $t=0$.
There are also some more subtle effects of the higher magnetic field that we will discuss
below in conjunction with our discussion of the time-dependent probability density of 
the scattered states in the wire.

\begin{figure}[htbq]
      \includegraphics*[width=0.42\textwidth,angle=0,viewport=10 10 345 212]{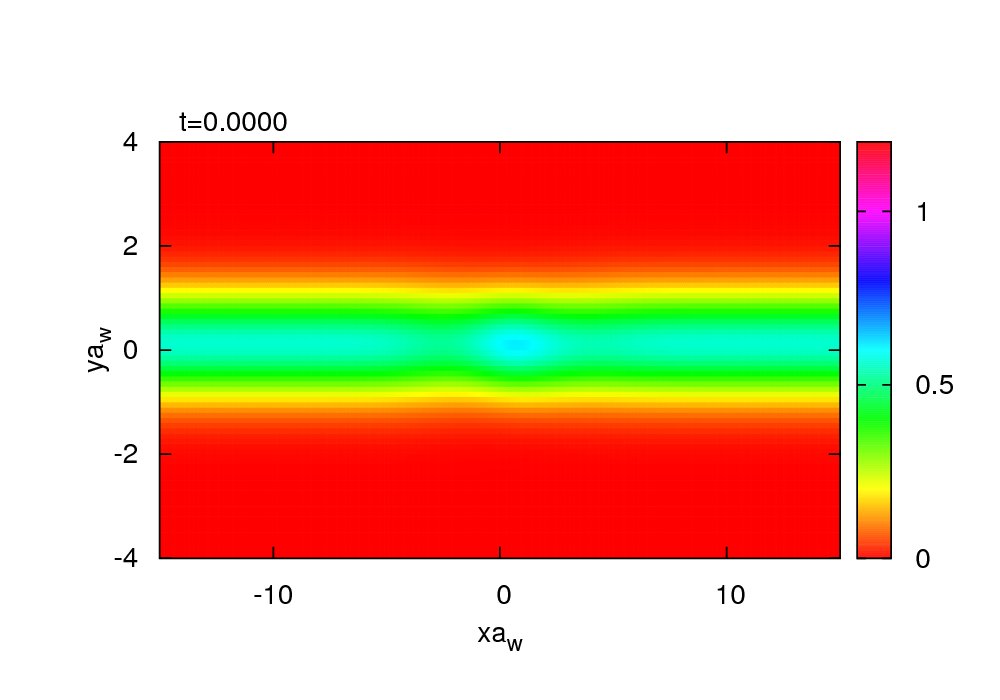}
      \includegraphics*[width=0.42\textwidth,angle=0,viewport=10 10 345 212]{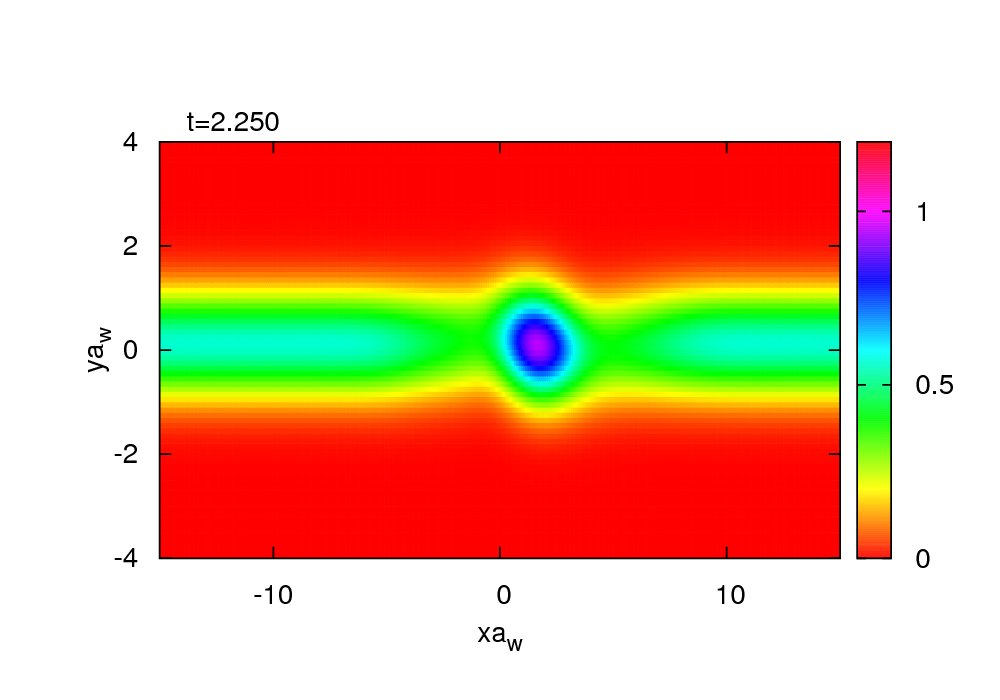}
      \includegraphics*[width=0.42\textwidth,angle=0,viewport=10 10 345 212]{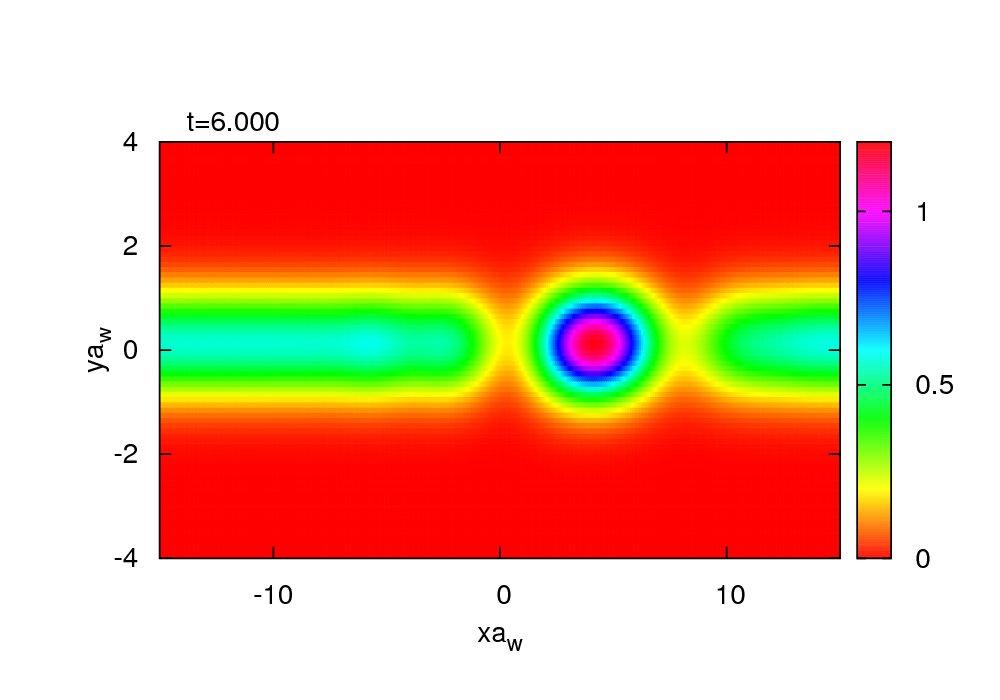}
      \includegraphics*[width=0.42\textwidth,angle=0,viewport=10 10 345 212]{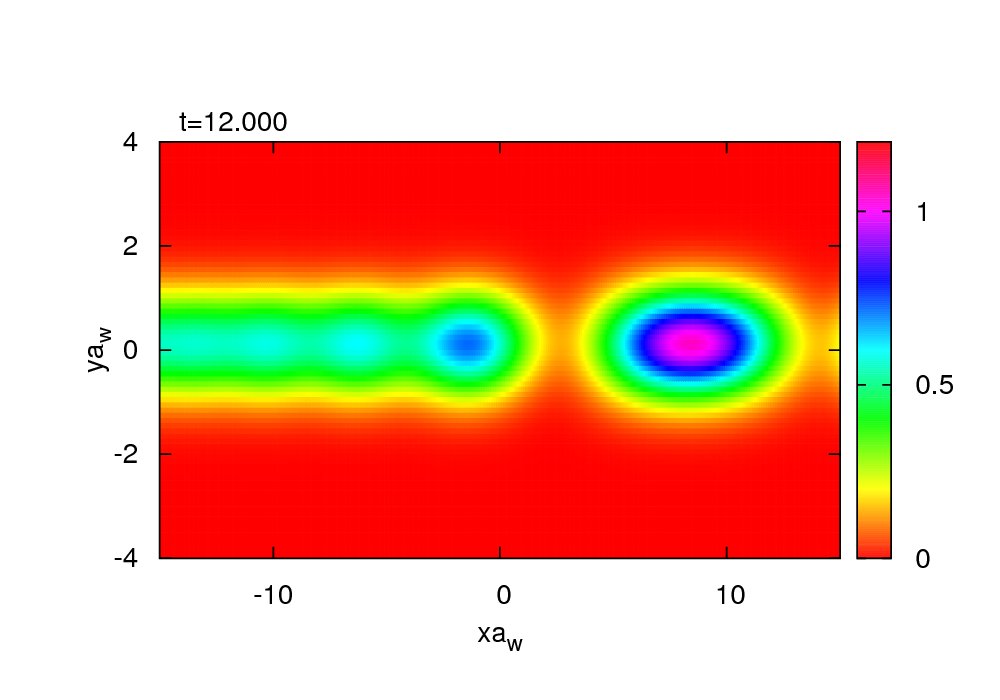}
      \caption{(Color online) The probability density $|\Psi ({\mathbf r},t)|^2$ for $k_na_w=0.724$
               corresponding to energy of the in-state $E=0.754\hbar\Omega_w$,
               $\beta =1\times 10^{-4}$ nm$^{-2}$, $B=0.1$ T, $V_0=-1.0$ meV,
               $\Omega =0.2\Omega_w$, and $\gamma = 1.0\Omega_w^{-2}$.}
      \label{Wft_B01_b01_kn08_m}
\end{figure}

As we discussed above the net current through our quantum wire will be carried by the
scattering state caused by the in-state with $E=\mu_l$. In order to gain more information
about the effects of the flashing on and off of the time-dependent scattering potential
$V_{\mathrm{sc}}$ in the wire than the current allows us we turn to analysing the
time-dependent probabilities for these important scattering states.
We start by looking at the probability density $|\Psi ({\mathbf r},t)|^2$ for the
incoming energy $E=0.754\hbar\Omega_w$, corresponding to the dimensionless momentum
$k_na_w=0.724$, for the formation of a broad well in the low magnetic field $B=0.1$
in Fig.\ \ref{Wft_B01_b01_kn08_m}.
Let us just restate that the flashed on and off temporal and spatial Gaussian well 
reaches its maximum at $t=0$ and retains more than half its depth only  
in the interval $t\Omega_w\in [-1,+1]$. It is thus interesting to notice the delayed
action on the probability reflected by the fact that at $t=0$ only a small variation
is visible. It is not until the well is almost totally turned off again that we
can see strong effects in terms of a wave packet or pulse, that then slowly propagates
in the system to the right. This wave packet representing a quasi-bound state 
that has temporarily formed and is then being released into the wire is the
cause of the current modulation that we saw above in Fig.\ \ref{DJB01m}. 
Just as the current Figure indicated the speed of the wave packet is determined by the
momentum of the incoming wave, and its spreading demonstrates the bandwidth of
the momentum or energy components that where assembled into it by inelastic
scattering processes. We also notice the delayed formation of a wave 
pattern to the left of the wave packet displaying the interference of the
incoming and the reflected waves. 
\begin{figure}[htbq]
      \includegraphics*[width=0.42\textwidth,angle=0,viewport=10 10 345 212]{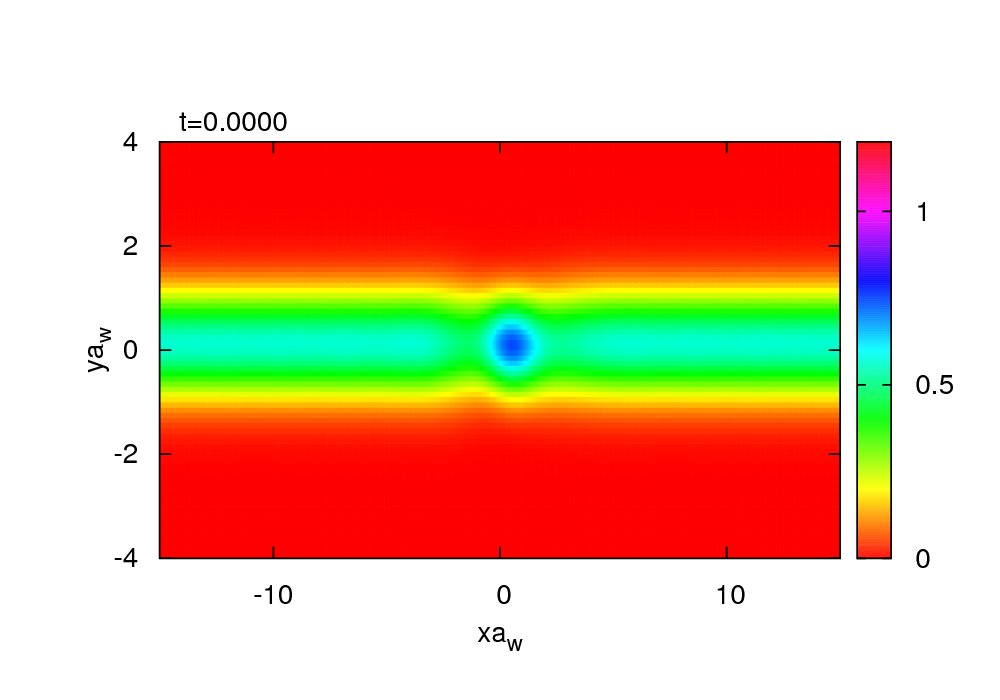}
      \includegraphics*[width=0.42\textwidth,angle=0,viewport=10 10 345 212]{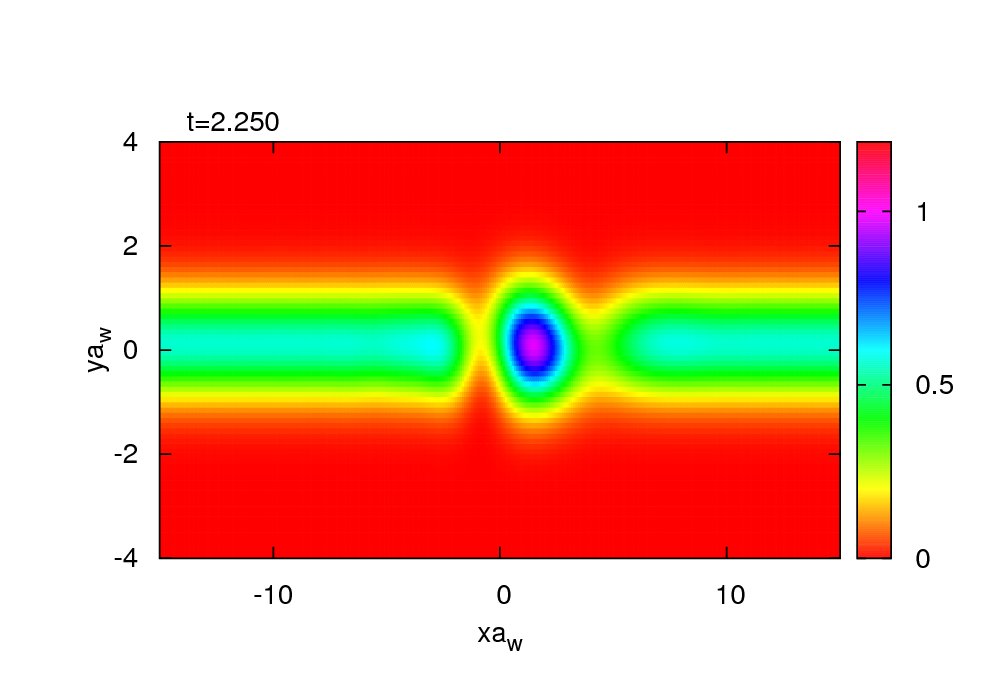}
      \includegraphics*[width=0.42\textwidth,angle=0,viewport=10 10 345 212]{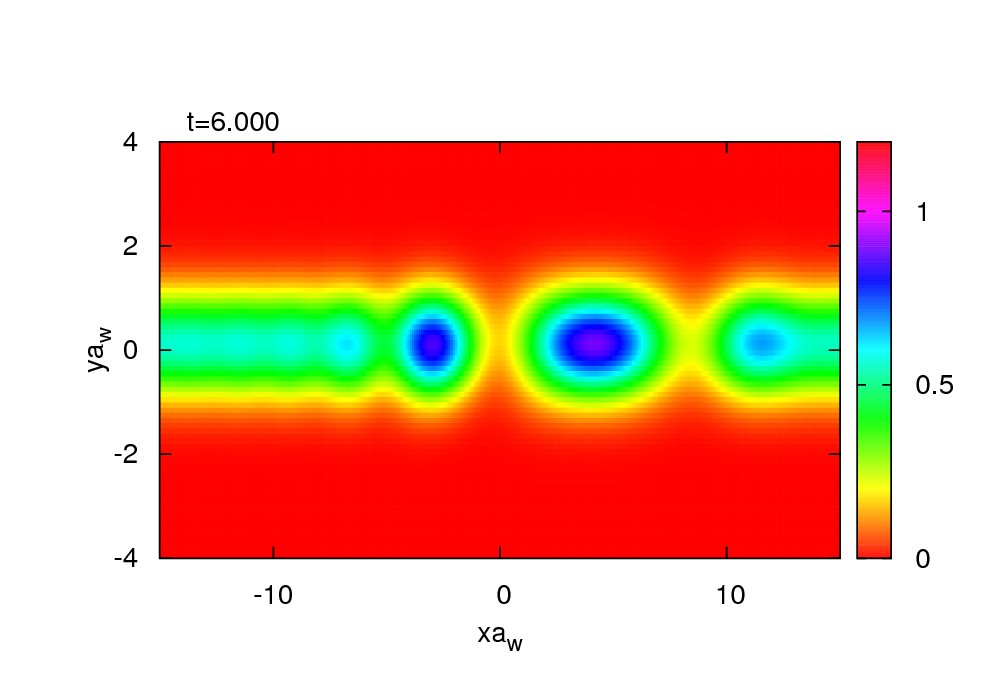}
      \includegraphics*[width=0.42\textwidth,angle=0,viewport=10 10 345 212]{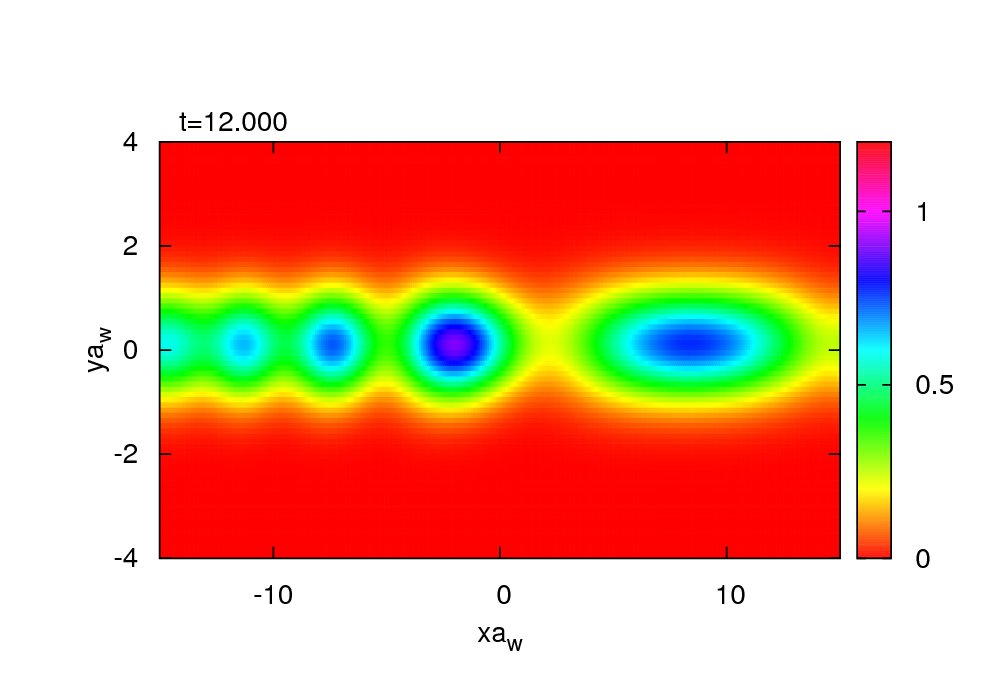}
      \caption{(Color online) The probability density $|\Psi ({\mathbf r},t)|^2$ for $k_na_w=0.724$
               corresponding to energy of the in-state $E=0.754\hbar\Omega_w$,
               $\beta =4\times 10^{-4}$ nm$^{-2}$, $B=0.1$ T, $V_0=-1.0$ meV,
               $\Omega =0.2\Omega_w$, and $\gamma = 1.0\Omega_w^{-2}$.}
      \label{Wft_B01_b04_kn08_m}
\end{figure}

The corresponding results for the flashing on and off for the narrower well (Fig.\ \ref{Vsc_D}(b))
are presented in Fig.\ \ref{Wft_B01_b04_kn08_m}. Here again we notice the formation of a 
wave packet, this time more confined, and correspondingly there is a stronger interference
visible between the incoming and the reflected wave. We will not show any corresponding 
results for the formation of a wave packet, or the releasing of a quasi-bound state for
higher magnetic fields here, but for a higher magnetic field the incoming wave is shifted
away from the center of the wire and thus the collision with the scattering potential
will not be totally a head-on collision. This off-center effect causes contributions from
higher subbands to be included in the wave packet resulting in a slow oscillation 
perpendicular to the wire. The magnetic field couples the motion in the $x$- and
$y$-directions such that these oscillations together with the propagation of the
packet to the right results in oscillating shape changes to the packet. One might
say that it looks like the wave packet for a higher magnetic field wobbles along the
wire. In an interacting many-electron system (that we are not exploring here) such
motion would inevitably couple to the plasmonic degrees of freedom. This wobbling effect is 
not easily visible in the Figures of the current presented above.

\begin{figure}[htbq]
      \includegraphics*[width=0.42\textwidth,angle=0,viewport=10 10 345 212]{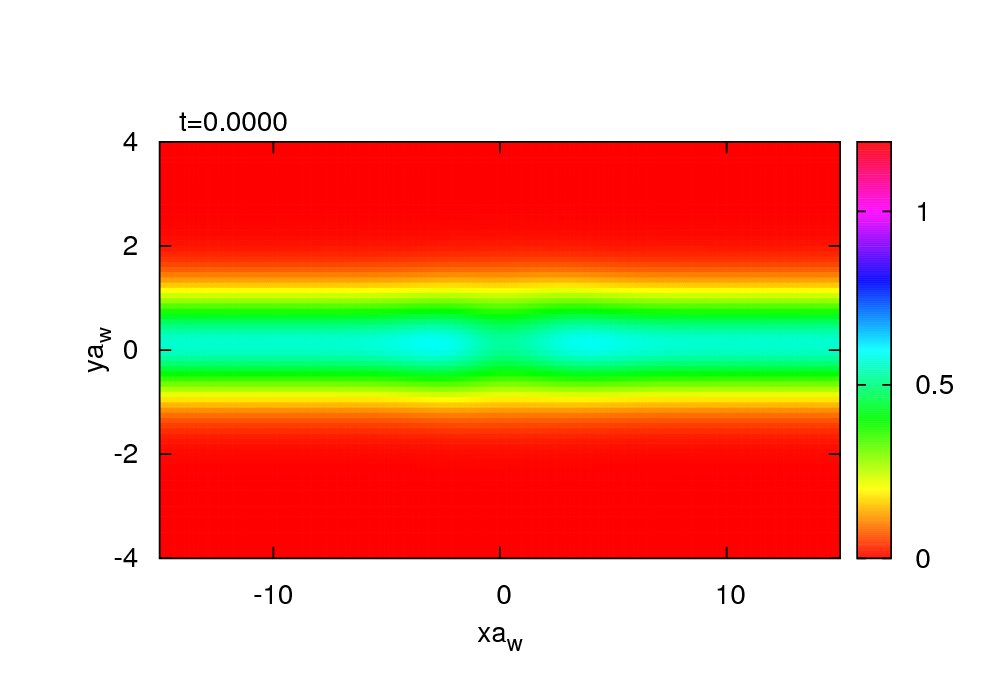}
      \includegraphics*[width=0.42\textwidth,angle=0,viewport=10 10 345 212]{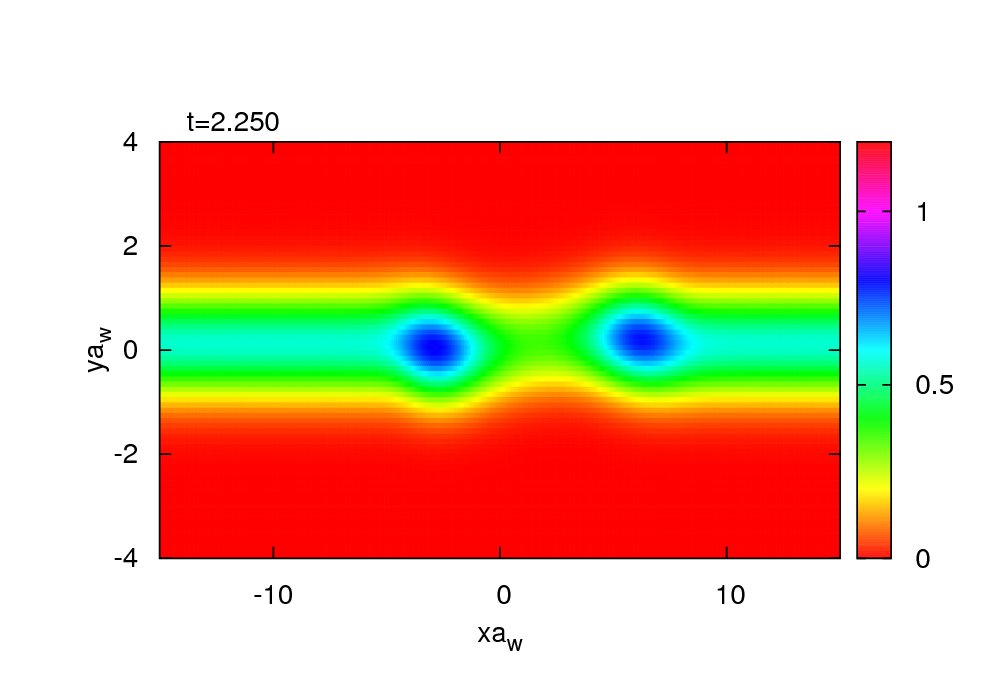}
      \includegraphics*[width=0.42\textwidth,angle=0,viewport=10 10 345 212]{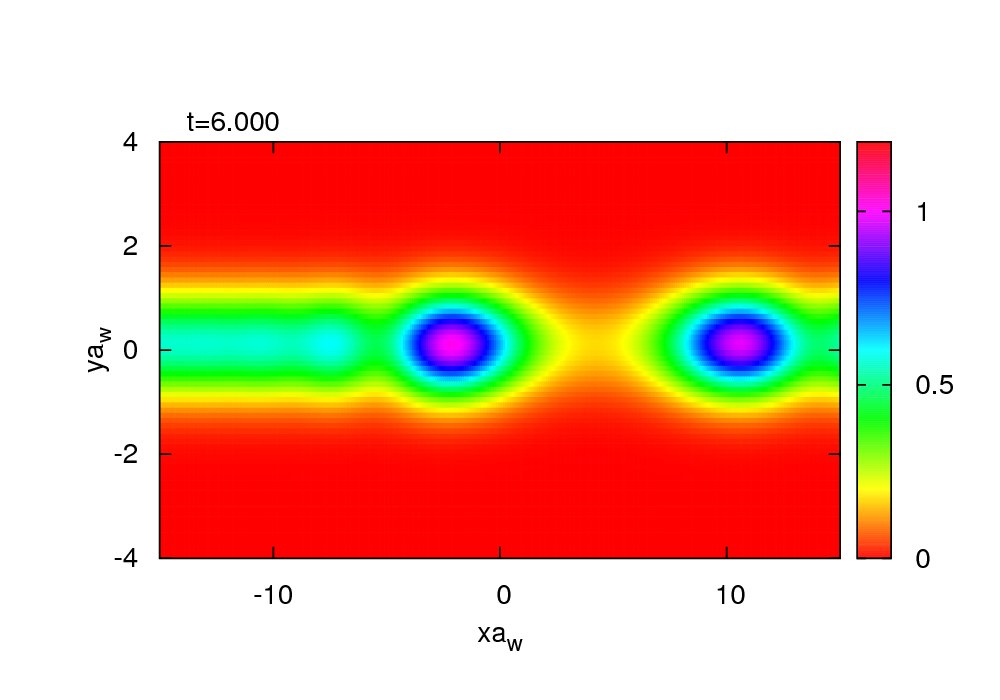}
      \includegraphics*[width=0.42\textwidth,angle=0,viewport=10 10 345 212]{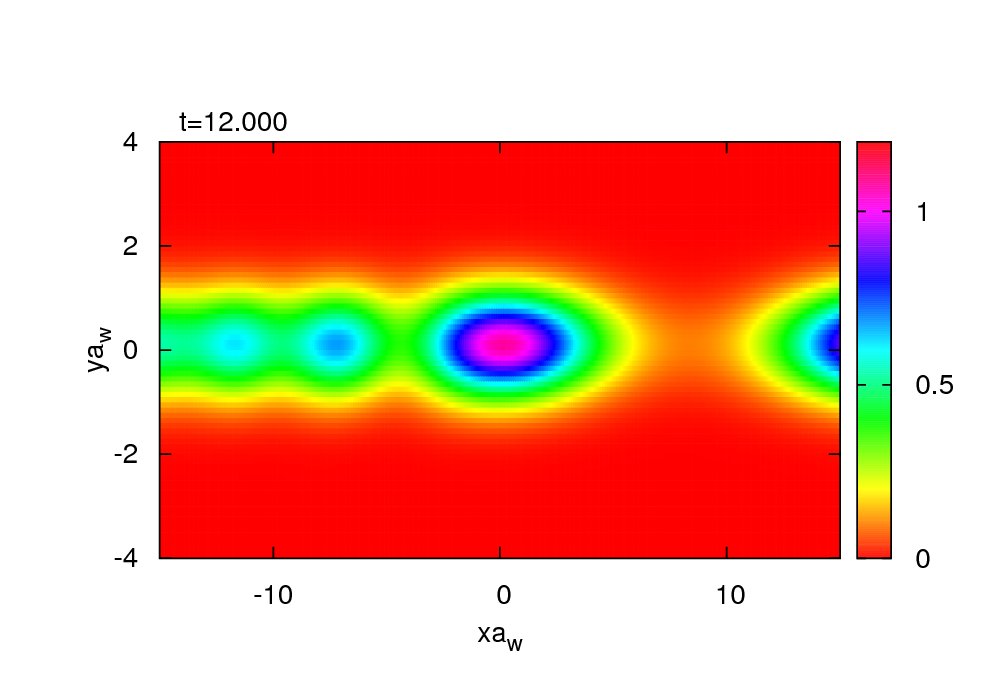}
      \caption{(Color online) The probability density $|\Psi ({\mathbf r},t)|^2$ for $k_na_w=0.724$
               corresponding to energy of the in-state $E=0.754\hbar\Omega_w$,
               $\beta =1\times 10^{-4}$ nm$^{-2}$, $B=0.1$ T, $V_0=+1.0$ meV,
               $\Omega =0.2\Omega_w$, and $\gamma = 1.0\Omega_w^{-2}$.}
      \label{Wft_B01_b01_kn08_p}
\end{figure}

\begin{figure}[htbq]
      \includegraphics*[width=0.42\textwidth,angle=0,viewport=10 10 345 212]{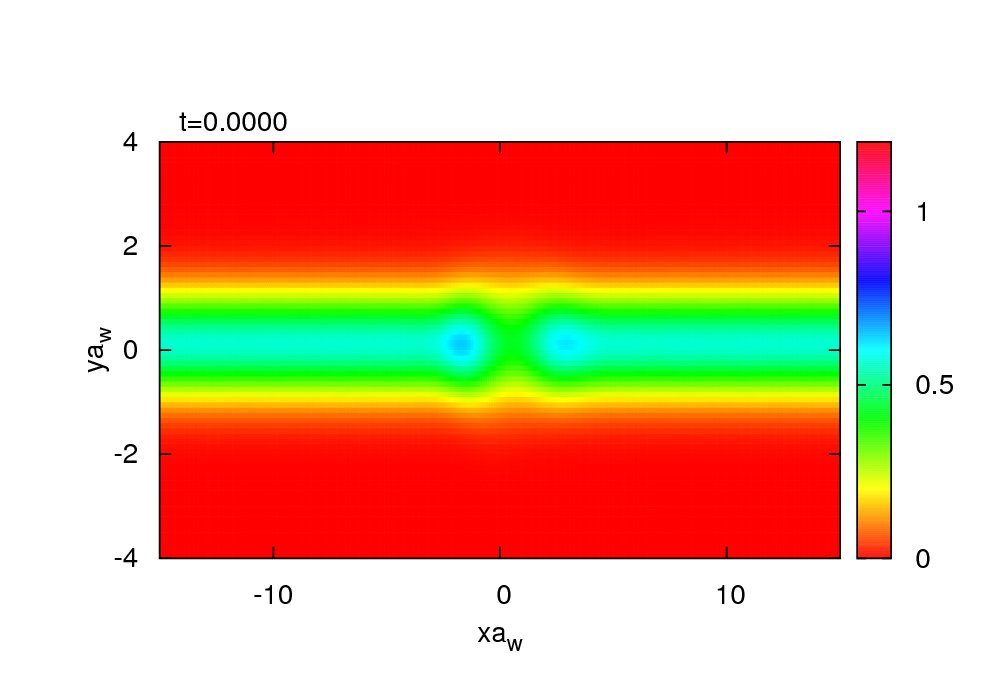}
      \includegraphics*[width=0.42\textwidth,angle=0,viewport=10 10 345 212]{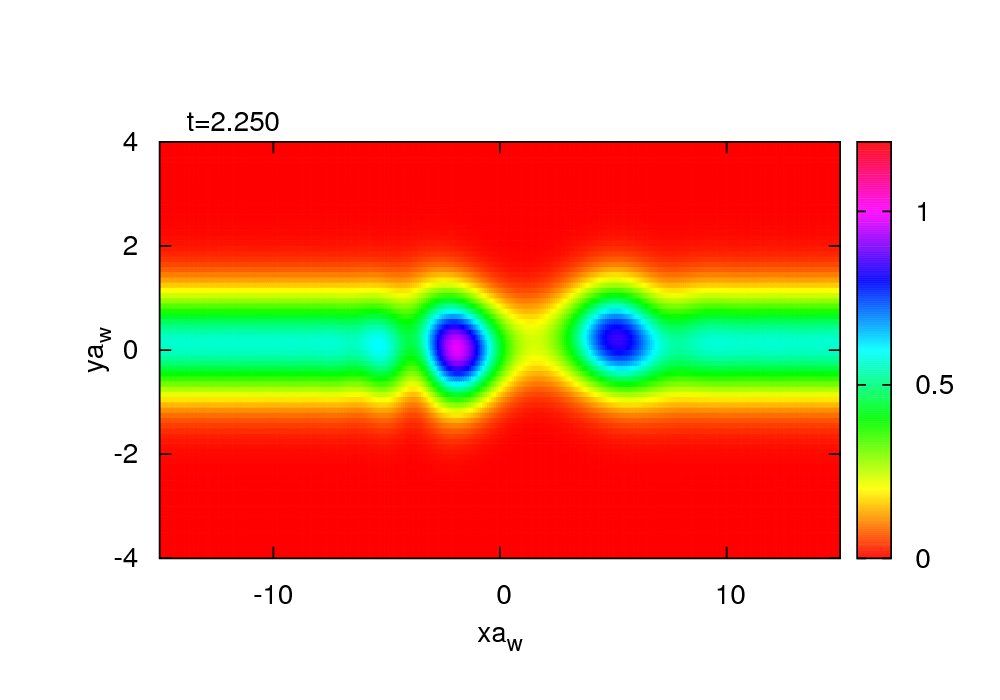}
      \includegraphics*[width=0.42\textwidth,angle=0,viewport=10 10 345 212]{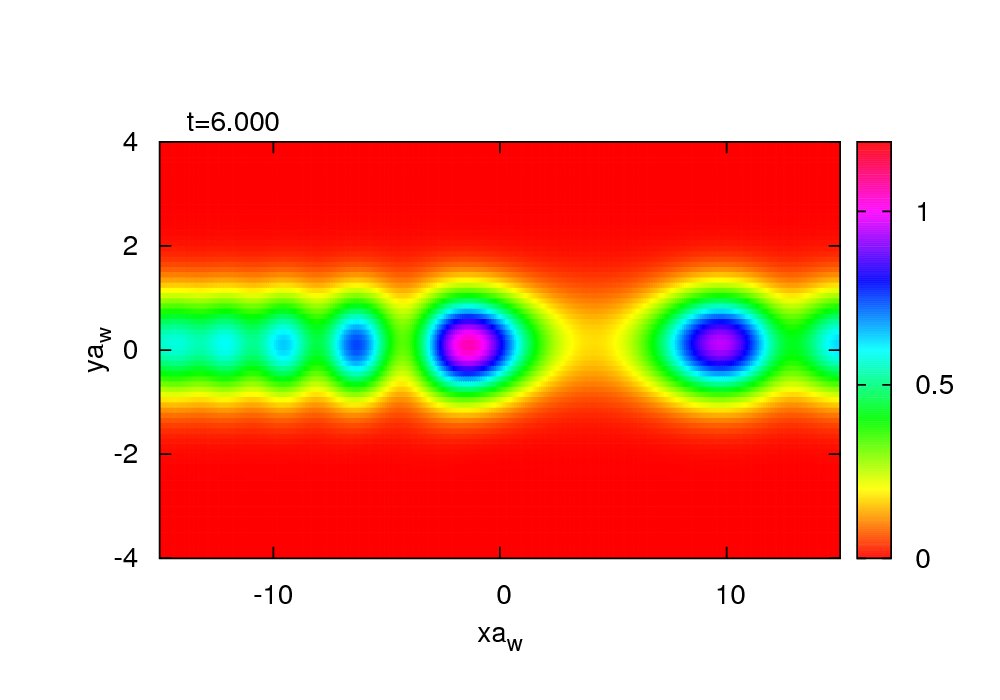}
      \includegraphics*[width=0.42\textwidth,angle=0,viewport=10 10 345 212]{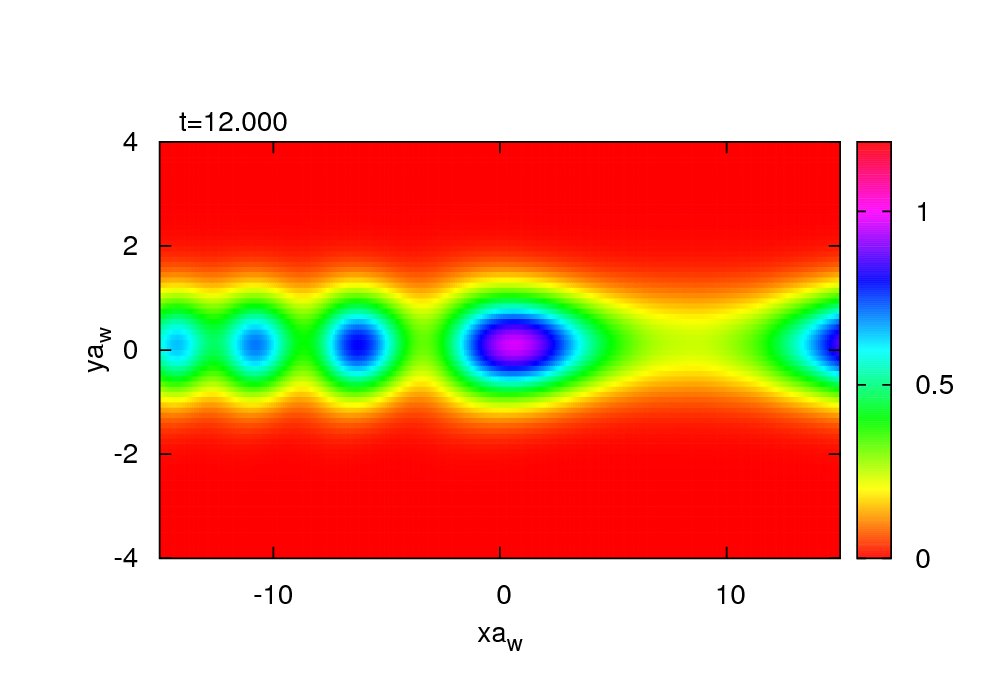}
      \caption{(Color online) The probability density $|\Psi ({\mathbf r},t)|^2$ for $k_na_w=0.724$
               corresponding to energy of the in-state $E=0.754\hbar\Omega_w$,
               $\beta =4\times 10^{-4}$ nm$^{-2}$, $B=0.1$ T, $V_0=+1.0$ meV,
               $\Omega =0.2\Omega_w$, and $\gamma = 1.0\Omega_w^{-2}$.}
      \label{Wft_B01_b04_kn08_p}
\end{figure}

The flashing on and off for the constriction leads to the formation of a saddle-shaped
wave packet as is seen in Fig.\ \ref{Wft_B01_b01_kn08_p} for the longer version
(Fig.\ \ref{Vsc_H}(a)), and in Fig.\ \ref{Wft_B01_b04_kn08_p} for the
shorter version (Fig.\ \ref{Vsc_H}(b)). 
As expected, this saddle-shaped wave packet has a longer extension in the case of the
longer constriction, and as noted above for the current, there is a considerable 
backscattering creating a clear interference pattern to the left of the scattering
region. Again this is understandably caused by the ``squeezing'' out action of the
constriction formation.

Just like for the single-hump wave packet at higher magnetic field, the saddle-shaped packet is seen to wobble 
along the wire for a higher magnetic field.

%
\section{Summary}
We have taken the initial steps to use a time-dependent Lippmann-Schwinger 
scattering formalism to explore the interplay of geometrical and magnetic field
effects on the transport of an electron with definite energy through a quantum
wire with an embedded scattering potential that is smoothly flashed on and off.
The selection of the spatial part of the scattering potential allows us to mimic
a temporary constriction or a broadening being introduced into the wire.
We have observed how this temporary change in the wire geometry can 
modulate the nonlocalized initial electron state in the wire to form
localized wave packets through inelastic scattering
processes in the wire. Another view on the formation of the wave packets
is that they represent quasi-bound states that form momentairily and are then
again released as the scattering potential vanishes. In support of this view
is the delay observed in the formation of the packet, it only clearly appears
as the potential has almost totally disappeared. 

The magnetic field through the Lorentz force causes the incoming wave to 
glance of the scattering potential instead of colliding with it head-on.
This off-centering of the scattering process mixes components of higher
subbands into the wave packet making it to wobble along the wire.

As expected for a wave system, the formation of a wave packet is accompanied
by an interference pattern being formed on the incoming side of the 
scattering center displaying the coexistence of the in-wave and the
reflected wave there. 

We note that a comparison between the results for a static systems and
our present time-dependent system is not simple. In the static systems
the reflectance of an incoming wave might be strong in some range of
the incoming energy and thus the conductance in this range would be low.
In the dynamical case we are studying here the potential only forms
for a short time and it might lead to a strong short-lived redistribution 
of the electron probability in the scattering range that will not necessarily 
cause strong reflection.
In addition, we should also state that a direct comparison to systems with
a time-harmonic scattering center is not appropriate since the short-lived
time-dependent scattering center in our model does not lead to the formation
of clear sidebands, common is though the inelastic character brought to the
collision by the time-dependent scattering potential. 
The Lippmann-Schwinger approach has also been used
to describe the scattering process in time-periodic atomic 
systems.\cite{Millack90:1693,Peskin94:3712} 

We stress again the single-electron character of our approach, and the
fact that it prevents us to talk about charge accumulation or depletion in
the scattering region. We can only discuss the change in the probability 
density of the electron as any reference to charge implicitly points out
the absence of the Coulomb interaction in our present model 


%
%
\begin{acknowledgments}
      The authors acknowledge financial support from the Research
      and Instruments Funds of the Icelandic State,
      the Research Fund of the University of Iceland, the
      Icelandic Science and Technology Research Programme for
      Postgenomic Biomedicine, Nanoscience and Nanotechnology, the
      National Science Council of Taiwan, and the National Center
      for Theoretical Sciences, Tsing Hua University, Hsinchu, Taiwan.
      C.S.T.\ is grateful to the computational facilities supported
      by the National Center for High-Performance Computing in Taiwan
      and the University of Iceland. The authors acknowledge technical
      assitance from Cai-Jhao Fan Jiang.
\end{acknowledgments}

%
%
\bibliographystyle{apsrev}

%
%
%
\end{document}